\documentclass[reqno]{amsart}
\usepackage[normalem]{ulem}
\usepackage[magyar]{babel}
\usepackage{amssymb,amstext,amsmath}
\usepackage{graphicx}
\usepackage{t1enc}              
\usepackage[cp852]{inputenc}    
\newcommand\re[1]{(\ref{#1})}

\begin{document}

\newcommand\al{\alpha}
\newcommand\ep{\epsilon}
\newcommand\epp{^p\!\!\epsilon}
\newcommand\epe{^e\!\!\epsilon}
\newcommand{\pd}[2]{\frac{\partial #1}{\partial #2}}
\newcommand{\pdt}[3]{\frac{\partial^2 #1}{\partial #2 \partial #3}}
\newcommand{\mg}[1]{\tilde\partial_{#1}}
\newcommand{\nev}[1]{#1} 

\def\SSS{\hbox{\boldmath{$\cS$}}}
\def\SSSS{\hbox{\boldmath{$\sS \cS$}}}
\def\Sh{\mathbf{h}}
\def\SSC{\hbox{\boldmath{$\cC$}}}
\def\SSCC{\hbox{\boldmath{$\sS \cC$}}}
\def\FFA{\tens{A}}

\title{A k\'epl\'ekenys\'egtan termodinamikai alapjair\'ol}
\author{Ván Péter}
\address{KFKI, RMKI, Elméleti Fizika Főosztály, Budapest,
BME, Energetikai Gépek és Rendszerek Tanszék, Budapest és
Montavid Termodinamikai Kutatócsoport, Budapest
}
\date{2010. január}
\maketitle

\section{Bevezetés}

Ebben az írásban a képlékeny és a reológiai, (hibás szóhasználattal 
viszkoelaszto-plasz\-tikus)
anyagmodellek kapcsolatát tárgyaljuk a  nemegyensúlyi termodinamika alapján. Annak
érdekében, hogy mondanivalónk könnyebben érthető és áttekinthetőbb legyen röviden
összefoglaljuk a legfontosabb fogalmakat\ a klasszikus rugalmasságtanból,
illetve a klasszikus és a hagyományos termodinamikai hátterű  
képlékenységtanból.

Többféle képlékenységi elmélet áttekintése után részletesen vizsgáljuk a \nev{Ziegler}-től
eredő klasszikus termodinamikai képlékenységelméletet, amely a legelterjedtebb és 
számos szempontból a legjobbnak tekinthető. Itt  a képlékeny deformáció egy
speciális  
belső változó, ami csökkenti a termosztatikai
feszültséget, vagyis az egyensúlyi állapotban lévő közeg feszültségét. A képlékenységi 
feltételt az \nev{Onsager}-féle vezetési együtt\-hatóknak a belső változó sebességének 
abszolút értékétől való függése hordozza. 

Ahogy a csúszási súrlódás kontinuumokra történő általánosítása vezet a viszkozitáshoz,
és speciálisan izotrop folyadékok esetén a \nev{Navier-Stokes}-egyenlethez, illetve
szilárd testek esetén a különféle viszkoelasztikus elméletek alapegyenleteihez,
ugyanúgy kaphatjuk a tapadási súrlódás kontinuummechanikai általásnosításaként
a képlékenység különféle elméleteit. Éppen ezért a termodinamikai elmélet
ismertetése előtt --- és a klasszikus és termodinamikai képlékenység különbségének
teljesebb megvilágítása céljából --- kitérünk a súrlódás és csillapítás termodinamikai
leírásaira.  

A termodinamikai képlékenységet teljesen a nemegyensúlyi
termodinamika fogalomrendszerén belül tárgyaljuk, eltérően a szokott, a mechanikában
kialakult fogalmakra alapozott tárgyalástól. A nemegyensúlyi termodinamika
a  képlékeny alakváltozásokkal járó folyamatok időbeli lefolyásának
leírását teszi lehetővé. A  képlékeny és a rugalmas deformáció időbeli változása
kúszási és feszültségrelaxációs jelenségekkel együtt leírható. Homogén testek
példáján --- közönséges diferenciálegyenletek megoldásával --- mutatjuk meg a disszipatív hatások numerikus regularizáló hatását.

\section{A képlékenységi elméletek rövid összefoglalása}

\subsection{A klasszikus képlékenységelmélet} 

A  kép\-lé\-kenység elméleteiben a teljes alakváltozást rugalmas és képlékeny részre
szokták oszatni. A nagy alakváltozásos elméletekben a teljes alakváltozást
a rugalmas és   a képlékeny alakváltozás szorzataként, 
vagy összegeként állítják elő.
A kis deformációs elméletekben mindkét feltevés a deformációk  összegződésére vezet.
Mi a továbbiakban a kis deformációs elmélettel foglalkozunk, ezért a teljes \(\ep_{ij}\) deformációt (strain) 
\(\ep_e^{ij} \) rugalmas és \(\ep_p^{ij} \) képlékeny 
komponensek összegére bont\-juk:
\begin{equation}
\ep^{ij} =\ep_e^{ij}+\ep_p^{ij}, 
\label{addpla}\end{equation}
ahol mindkét deformáció  másodrendű tenzor.  A későbbiekben megmutatjuk, hogy
egy termodinamikai belső változó milyen feltételekkel értelmezhető képlékeny
deformációként. Itt és a továbbiakban a tenzorokat
alsó és felső indexekkel jelöljük, és az egy szorzaton
belül ismételt indexek összegzést jelentenek,  az \nev{Einstein}-féle összegzési 
szabálynak megfelelően. Általában ügyelünk az alsó és felső indexek megfelelő
használatára is.

A rugalmas feszültséget a
termosztatikából ismert módon  termodinamikai poten\-ciálfüggvényből származtatjuk.
Ez a termodinamikai alapállás megfelel a mechanikában  hiperrugalmasságként
ismert elméletcsaládnak. Mivel az entrópia és rajta keresztül a termodinamikai potenciálok
létezése a második főtétel részeként természettörvény,, ezért termodinamikai szempontból 
az ilyen elméletek kitüntetettek. Mechanikai
elméletekben a leggyakrabban használt termodinamikai potenciál az \(F\) szabadenergia-függvény
(illetve sűrűség).
Ennek segítségével 
\begin{equation}
 \sigma_{ij} = \pd{F}{\ep_e^{ij}}.
\label{stress}\end{equation}

 A vonatkozó \nev{Gibbs}-reláció \(dF= SdT+\sigma_{ij} d \ep^{ij}\), ahol \(S\) az
 entrópia, \(\sigma_{ij}\) a feszültség és \(T\) a hőmérséklet.
Ideálisan rugalmas anyagok esetén 
\begin{equation}
\sigma_{kl} = C_{ijkl}\ep_e^{ij},
\label{idrug}\end{equation}
ahol a $C_{ijkl}$ negyedrendű
rugalmassági tenzor állandó, továbbá a szabadenergiából történő származtathatóság
követelménye miatt \(C_{ijkl}=C_{klij}\). Ideálisan rugalmas izotróp anyag esetén 
\(C_{ijkl}= \lambda\delta_{ij}\delta_{kl}+\nu \delta_{ik}\delta_{jl}\) és
a szabadenergia
\begin{equation}
F(\ep_e^{ij}) = \frac{\lambda}{2} ((\ep_e)_{i}^i)^2+ \mu (\ep_e)_{ij}\ep_e^{ij}.
\end{equation}

Itt \(\lambda\) és \(\mu\) a \nev{Lamé}-állandók, \(\delta_{ij}\) pedig a másodrendű
egységtenzor. A képlékenység határát egy ún. folyási felület 
(yield surface) határozza meg a fe\-szültségtérben.
Ezt a \emph{folyásfüggvény} segítségével adják meg, a $0$ értékéhez 
tartozó szintfelülettel: 
$$
f(\sigma_{ij},...)=0.
$$ 
Az
\(f\) függ\-vény más fizikai mennyiségektől  is függhet. Képlékeny
alakváltozás akkor következik be, ha a feszültség eléri a  folyásfelületet.
Képlékeny alakváltozás közben ott is marad.
A folyásfelületen "belül", ahol \(f(\sigma_{ij},...)<0\), az anyag rugalmasan viselkedik,
a folyási felületen kívüli rész, ahol \(f(\sigma_{ij},...)>0\), nem érhető el. A folyási 
függvény nincs egyértelműen meghatározva, egyetlen szintfelülete lényeges, ezért a   
fenti tulajdonságokkal minden monoton függvénye rendelkezik. Többnyire implicit 
módon, de feltételezik továbbá, hogy a folyási felület időben állandó, azaz 
\begin{equation}
\dot f =0.
\label{fdot}\end{equation}

Ez tulajdonképpen rögzíti a folyási határ valamelyik változójának evolúciós egyenletét.

A képlékeny deformáció változását a \emph{folyási törvény} (flow rule) adja meg,
amit általában a \(g(\sigma_{ij},...)\) \emph{képlékeny potenciál} segítségével írnak fel,
feltételezve, hogy a képlékeny deformáció változása merőleges a képlékeny
potenciál szintfelületeire:
\begin{equation}
(\dot\ep^p)^{ij} = \Lambda \pd{g}{\sigma_{ij}}, 
\label{flowrule}\end{equation}
ahol \(\Lambda\)  pozitív skalár képlékeny szorzó. Ha \(f\equiv g\), akkor \textit{kapcsolt
vagy asszociatív\ folyás}ról (associated flow), illetve normalitásról beszélünk. A képlékeny
potenciált sokszor egy segédváltozóval lenullázzák a folyásfelület azon
pontjaiban, ahol a képlékeny folyás történik, felhasználva a szintfelületes
meghatározás miatti határozatlanságot. 

A képlékeny deformáció mértékének meghatározására változatos elképzelések
vannak. Leggyakrabban feltételezik, hogy a folyásfüggvény egy 
\(\xi\) \emph{keményedési (hardening) paramétertől } függ. Ha a keményedési
paraméter egyedül a kép\-lékeny deformáció függvénye, akkor \(\xi\) 
kiküszöbölhető a folyásfüggvényből, és \textit{deformációs keményedés}ről
(strain hardening) beszélnek.  Máskor ugyan a  keményedési paraméter
nem küszöbölhető ki, de rá vonatkozóan fejlődési egyenleteket írhatunk fel,
például \(\dot \xi = x(\sigma_{ij},(\ep^p)^{ij},\xi,(\dot\ep^p)^{ij}) \) formában,
ahol ay egyenlet jobb oldalán álló \(x\) függvény mutatja a fejlődési egyenlet
szokásos változóit.
A keményedési paraméter legtöbbször skalár. Ha a keményedés változását éppen a képlékeny teljesítmény okozza, azaz 
\(x= \dot W^p = \sigma_{ij} (\dot \ep_p)^{ij}\), akkor beszélünk \emph{munkakeményedésről}
(work hardening). 

Mindezek a feltételek együtt megadják a képlékeny szorzót és megadják a rugalmasági
paraméterek megváltozott, képlékeny tartományban érvényes értékeit is. Nézzünk
erre két példát. 

\emph{a) Ideális és  nem kapcsolt képlékenység}
esetén, amikor a folyási függvény csak a feszültségtől függ, azaz \(f=f(\sigma_{ij})\):
\begin{equation}
\dot f = \pd{f}{\sigma_{ij}} \dot \sigma_{ij} = 
  \pd{f}{\sigma_{_{ij}}}C_{ijkl}\dot \ep_e^{kl}=
  \pd{f}{\sigma_{ij}}C_{ijkl}
    \left(\dot \ep^{kl} - \lambda \pd{g}{\sigma_{kl}}\right).
\end{equation}
 
Ahol az előbb már bevezetett negyedrendű
rugalmassági tenzort az általános nemlineáris esetben a szabadenergia második
deriváltjaként  
\(C_{ijkl}=\pdt{F}{\ep_e^{ij}}{\ep_e^{kl}}\) módon értelmezzük. Ebből \re{fdot} alapján:
\begin{equation}
\lambda = \frac{\pd{f}{\sigma_{ij}}C_{ijkl}\dot \ep^{kl}}{
  \pd{f}{\sigma_{mn}}C_{mnrs}\pd{g}{\sigma_{rs}}}.
\end{equation}

Ezek után a feszültség megváltozása
\begin{equation}
 \dot \sigma_{ij} = C_{ijkl} \dot \ep^{kl} -
     \left( \frac{\pd{f}{\sigma_{ab}}C_{abcd}\dot \ep^{cd}}{
  \pd{f}{\sigma_{mn}}C_{mnrs}\pd{g}{\sigma_{rs}}}\right)
  C_{ijkl}\pd{g}{\sigma_{kl}}.
\label{idplfin}\end{equation}

Kiemelve \(\dot\ep^{ij}\)-t leolvashatjuk a képlékeny viselkedés tartományában
érvényes módosított rugalmassági modulust: 
\begin{equation}
 \hat C_{ijkl} = C_{ijkl} -
     \left( \frac{\pd{f}{\sigma_{ab}}C_{abkl}}{
  \pd{f}{\sigma_{mn}}C_{mnrs}\pd{g}{\sigma_{rs}}}\right)
  C_{ijcd}\pd{g}{\sigma_{cd}}.
\label{idplrugmod}\end{equation}

\emph{Deformációs keményedés }esetén a folyási felület a feszültségnek is
függvénye, ezért írhatjuk, hogy
\begin{displaymath}
\dot f  =  \pd{f}{\sigma_{ij}} \dot \sigma_{ij} + 
        \pd{f}{\ep_p^{ij}}\dot \ep_p^{ij}
  =   \pd{f}{\sigma_{ij}} \dot \sigma_{ij} + 
  \lambda\pd{f}{\ep_p^{ij}} \pd{g}{\sigma^{ij}}
  =0.
\end{displaymath}

Itt behelyettesítettük a \re{flowrule} folyási törvényt.
Ezután akár eljárhatunk hasonlóan is, mint az előbb, de esetleg kényelmesebb
lehet a a feszültségnövekményekre alapozva számolni. Azaz a képlékenységi szorzó meghatározására
a fenti egyenlőségből adódó következő formulát használjuk:\begin{equation}
\lambda = - \frac{  \pd{f}{\sigma_{ij}} \dot \sigma_{ij}}
{\pd{f}{\ep_p^{ij}} \pd{g}{\sigma_{ij}}}=
   \frac{  \pd{f}{\sigma_{ij}} \dot \sigma_{ij}}{h},
\end{equation}
ahol a \(h=- \pd{f}{\varepsilon_p^{ij}} \pd{g}{\sigma^{ij}}\) kombinációt
\emph{keményedési modulus}nak hívják.
Ezután a rugalmassági tenzor  helyett annak inverzét, a  \(c^{ijkl}\) merevségi tenzort
fogjuk használni, amelyre definíciójából következően igaz, hogy:
\begin{displaymath}
        \ep_e^{ij} = c^{ijkl} \sigma_{kl} .
\end{displaymath}
Természetesen a merevségi tenzor is származtatható potenciálból. Ennek segítségével írhatjuk,
hogy
\begin{equation}
 \dot \ep^{ij} = c^{ijkl} \dot \sigma_{kl} - 
        \frac{1}{h} \pd{g}{\sigma_{ij}} \pd{f}{\sigma_{kl}} \dot \sigma_{kl}.  
\label{idplno}\end{equation}

Ezért aztán a képlékeny tartományban érvényes merevségi tenzort könnyedén
kiolvashatjuk: 
\begin{equation}
 \hat c^{ijkl} = c^{ijkl} + \frac{\frac{1}{h} \pd{g}{\sigma_{ij}} \pd{g}
        {\sigma_{kl}}}{\pd{f}{\ep_p^{mn}} \pd{g}{\sigma_{mn}}} .
\label{idplmermod}\end{equation}

A klasszikus képlékenység elméletében csak erre van szükség, a feszültség és
a deformáció kapcsolatának úgynevezett \emph{növekményes} (incremental)
formáira, mint \re{idplfin}, vagy \re{idplno}.  A végeselem-prog\-ramoknak ennyi elég.  Annak
ellenére, hogy látszólag időderiváltak szerepelnek benne \re{idplfin}
vagy \re{idplno} csak nagyon korlátozott
feltételekkel vonatkoztatható valódi időbeli változásokra,  a rugalmassági
állandók csökkenését adja meg adott feszültségszint elérésekor.

Összefoglalva az eddigieket, egy klasszikus képlékenységelmélet feltételezi, hogy
\begin{enumerate}
\item A deformáció felbontható képlékeny és rugalmas komponensekre. \label{kl1}
\item A képlékeny viselkedés határát kritikus feszültségekkel jellemezhetjük, Ennek megfelelően
a feszültségtérben definiált folyásfüggvényt
egy szintfelületével adjuk meg, \(f(\sigma_{ij},...)=0\) módon. Ennek definíciója tartalmazza
a képlékeny deformáció irreverzibilitását, azaz azt a feltevést, hogy a képlékeny
deformáció egyúttal maradó deformáció, ha egyszer fellépett, akkor magától
nem csökken. \label{kl2}
\item A képlékenységi határ az \re{fdot} összefüggés szerint állandó. \label{kl3}
\item Létezik a \(g(\sigma_{ij},...)\) képlékeny potenciál. Azaz a képlékeny deformáció 
növekményeinek viszonyát a \re{flowrule} folyási szabály alapján - eléggé
speciális módon - jellemezhetjük (ez a \nev{Perzyna}-elmélet lényege). \label{kl4}  
\item Létezik valamilyen szabály  a képlékeny deformáció nagyságának meg\-ha\-tározásához
(pl. deformációs képlékenyedés, vagy a keményedés fej\-lődési egyenlete). \label{kl5}  
\end{enumerate}

Mindezek előtt, a \re{stress} összefüggés formájában adott a termosztatikai háttér 
- hiszen a szabadenergia létezése az entrópia bevezethetőségét feltételezi - az egyensúlyi feszültség és a rugalmas deformáció viszonyának megadására.

Termodinamikai szemmel vizsgálva a fenti feltevésrendszert,  érdekes, hogy a többi empirikusan megadandó függvény és közvetlen tapasztalati
sza\-bály mellett a \re{stress} összefüggést,
a szabadenergia,  azaz tulaj\-donképpen az entrópia létezésének feltevését
is gyengíteni szokták
valamilyen feszültség-deformáció függvény feltételezésével, annak potenciálból
történő származtathatósága nélkül. 
Erre a motivációt még \nev{Truesdell} és \nev{Noll} \cite{TruNol65b,Boj88b} adta a hiporugalmassági 
elképzelésükkel, akik így illesztették a reológiai jelenségeket a
mechanikához. A hiporugalmasság a feszültség növekményére vonatkozóan posztulál
függvénykapcsolatot \begin{equation}
\dot \sigma_{ij} = H_{ij}(\sigma_{ij}, \dot \ep_{ij}).
\label{hipo}\end{equation}
 Ennek mintájára jött létre \nev{Kolymbas} hipoképlékenységnek nevezett
elmélete. Ebben elhagyjuk a képlékeny potenciált is és a képlékeny deformáció 
meghatározására is egy  ilyen függvényt
keresünk (ami természetesen nem lineáris és legfeljebb az anyag szim\-metriái
szorítják meg) \cite{Kol00b}. A hiporugalmasságban és  hipoképlékenységben
a  termodinamikai követelményeket csak nagyon nehézkesen adhatjuk meg. A
továbbiakban látni
fogjuk, hogy a
\re{hipo}-hoz hasonló, a feszültség időderiváltját tartalmazó  egyenleteket viszont
a nemegyensúlyi termodinamika segítségével, matematikailag is következetesen
és könnyedén levezethetünk.
   
\subsection{A termodinamikai képlékenységelmélet}

A klasszikus képlékenységnek
a termodinamika második főtételéhez való viszonya nem tisztázott. Sok anyagra
nem tudjuk, hogy
 a folyási és képlékeny potenciálra
pontosan miféle követelményeket kellene még kikötnünk, hogy képlékenyedő anyagokkal
se lehessen másodfajú perpétum mobilét létrehozni, illetve a képlékeny anyagfüggvényeket
tartalmazó kontinuummechanikai modellek termodinamika egyensúlya aszimptotikusan
stabil legyen. A klasszikus képlékenység alapján még az sem világos, hogy
a mechanikai hiszterézis egyáltalán irreverzibilis jelenség-e. 
Számos  termodinamikai elmélet létezik különféle hiányosságokkal.

Az első jelentős termodinamikai elmélet, \nev{Rice} belső változós elképzelése \cite{Ric71a} 
a második főtételhez
köti, abból bizonyítja a normalitást.
Ez ma is az egykristály-képlékenység termodinamikai alapja, minden egyes diszlokációhoz
különböző belső változókat rendelve. 

A makroszkopikus képlékenység klasszikus  termodinamikai elmélete egyetlen
belső változóra - a képlékeny deformációra - alapuló  sajátos nemegyensúlyi
termodinamikai
elmélet. Az elmélet \nev{Hans Ziegler}től ered \cite{Zie81b}, és a francia iskola  
\cite{Mau92b,Mau99b} 
dolgozta ki (\nev{Duhem} műveiből is már kikövetkeztethető \cite{Maugin09o}.)  

A termodinamikai képlékenységelméletben a képlékeny deformációt termodinamikai bel\-ső
változónak tekintjük és fejlődési egyenletét az entrópiaprodukció egy részeként
azonosított disszipációs  függvényből származtatjuk. A vezetési egyenletek
a termodinamikai ára\-mokra bevezetett disszipációs potenciál formájában jelennek meg. 
A disszipációs potenciál egyben  képlékeny potenciál és folyási függvény is
--- a termodinamikai képlékenység alapkiépítésben kapcsolt elmélet. A legfontosabb
posztulátuma, hogy a disszipációs potenciál a termodinamikai áramoknak nem kvadratikus, 
hanem elsőrendű homogén függvénye
 (ideális képlékenység esetén). A termodinamikai
képlékenység természetes módon tartalmazza a viszkózus hatásokat is, az ideális
kép\-lékenység egyfajta szinguláris eset, a potenciálok differenciálhatóságát
sértő módon jelentkezik. Emiatt az ideális képlékenységet magába foglaló,
a klasszi\-kus képlékenység formalizmusát pontosan azonosító tárgyalása speciális 
matematikai 
esz\-közök bevezetését igényli (pl. \nev{Legendre-Fenchel} transzformáció).
A nem ideális - viszkoképlékeny - elmélet alapegyenletei az ideális képlékenységi
egyenletek egyfajta regularizációját eredményezik \cite{Gur00a}. 

 \nev{Ziegler} - elég nehezen követhető - érvelése és nevezetes ortogonalitási feltétele 
 termosztatikai 
kiindulóponton alapul \cite{Zie81b,ZieWeh87a}: 

Általában az entrópia változása reverzibilis és irreverzibilis részre osztható:
\begin{equation}
  dS = T d_r S + d_iS, \quad \text{ahol}, \quad d_iS \geq 0.
\label{kentr}\end{equation}

Tegyük fel, hogy az entrópia csak az \(U\) belső energiától és \(a_{k}\)
belső változóktól függ. Ekkor a \nev{Gibbs}-reláció a következő
formában írható:
\begin{equation}
  dU= T dS-A_k da_k = T d_rS -A_k da_k + Td_iS.  
\end{equation}
 Az utolsó tag - mint irreverzibilis járulék - folyamatsebességek függvénye és ezt tekinti
\nev{Ziegler} a \emph{disszipációs függvény}nek, azaz \(\Phi(U,\dot U, a_k,\dot a_{k}) = T
d_iS\). Ezután \nev{Ziegler} feltételezi, hogy 
\begin{enumerate}
\item \label{z1} az irreverzibilis járulék csak a belső változókhoz kötődik és ezért
\(Td_iS = F_kda_k\), ahol \(F_k\) általánosított disszipatív erőket jelöl,\label{termcond}
\item \label{z2} a disszipatív erők
párhuzamosak a disszipációs függvény növekedési irányával a nemegyensúlyi
állapottérben, azaz a következő kapcsolatban vannak 
\begin{equation}
  F_{k}= \nu\pd{\Phi}{\dot a_k}, 
\label{keplcond}\end{equation}
ahol \(\nu\) pozitív skalár értékű függvénye a \(\Phi\) változóinak.
\end{enumerate}

Ezek után - még a homogén rendszerekre vonatkozó termosztatikai keretek között - \nev{Ziegler}
kiaknázza \re{kentr} egyenlőtlenségét, a tulajdonképpeni entrópiaprodukciót
is, amely az eddigiek alapján 
$$F_k da_k \geq 0 $$
formában írható. Intuitív módon onsageri erőnek tekinti \(da_k\)-t és meghatározandó 
áramnak \(F_k\)-t,
és egyrészt megállapítja, hogy   
\begin{itemize}
\item 
a vezetési mátrix antiszimmetrikus része nem járul hozzá az entrópia\-produk\-cióhoz,
azaz a termodinamika - az ő szóhasználatával - nem mond semmit a "giroszkópikus
erőkről". 
\item nemdisszipatív esetben, azaz ha \(d_iS =0\), az \(F_{k}\) általánosított
disszipatív erők merőlegesek a \(da_k\) áramokra. Ez a híres \nev{Ziegler}-féle
\emph{ortogonalitási feltétel}, a klasszikus képlékenységtan termodinamikai
megalapozásának sokat vitatott sarokköve. Az ortogonalitási feltétel csak a nem disszipatív
esetben következménye a fenti gondolatmenetnek.
\end{itemize}
Az első \re{z1} feltevést \nev{Ziegler} semmivel sem próbálja indokolni,  annyira természetesnek
érzi. A másodikat, \re{z2}-t, később duális terekre vonatkozó (nem túl meggyőző) gondolatmenettel
támogatja meg (lásd \cite{ZieWeh87a}) illetve egy plauzibilisnek látszó variációs elvet, a 
Maximális
Entrópiaprodukció Elvét posztulálja helyette (vigyázat, ez nem azonos a \nev{Prigogine}-féle
Minimális Entrópiaprodukció Elvével, ami  vegyész és biológus körökben népszerű).   
Valójában úgy tűnik \nev{Ziegler} felismerte,
hogy ezzel a feltevéssel tudja a klasszikus képlékenységelméletek fogalomrendszerét
a nemegyensúlyi termodinamikával összekapcsolni, mert a disszipációs potenciál
ekkor természetes módon a folyási és a kép\-lékenységi függvényhez köthető.

Ziegler javaslatát azután   \nev{Maugin} (illetve a francia iskola) többféle irányban 
kiaknázza a mechanikai alapú elméletek
termodinamikai általánosításaival (töredezés, károsodás, anyagi sokaságok,
stb.) \cite{Mau93b,Mau99b,Mau99b1}. Ő mechanikai alapon - \nev{Hamilton-} típusú
variációs elvvel -
  próbálja indokolni Ziegler 
feltevéseit.  Fontos felismerése, hogy  az \nev{Onsager}-szimmetria
miatt a \nev{Ziegler} által még többféle változórendszerben meghatározott disszipációs függvénye
mögött felismeri  és azonosítja a
klasszikus \nev{Rayleigh}-féle disszipációs potenciálokkat.

Összefoglalva: a \nev{Ziegler}-féle termodinamikai képlékenységelmélet a klasszikus
kép\-lékenység feltételeit  egyszerűsíti az előző fejezet \ref{kl2}-\ref{kl5} pontjainak
alábbi módosításával:

\begin{itemize}
\item Disszipációs pontenciálként értelmezi a képlékenységi függvényt, és illeszti a 
nemegyensúlyi termodinamika elméletéhez. 
\item Lehetővé teszi  a képlékeny alakváltozásra vonatkozó fejlődési egyenlet
termodinamikai (vagy variációs mechanikai) alapon történő származtatását
\cite{MauMus94a1,Mau90a1}.
\end{itemize}

A fenti feltevésrendszer világosan megalapozza, illetve termodinamikai elmélet
keretei közé illeszti a kapcsolt képlékenységet,
és lehetőséget teremt az egyéb termodinamikai kölcsönhatásokkal együttes
tárgyalására. Ilyenek például a viszkózus hatásokat is figyelembe vevő viszkoplaszticitás,
károsodás, stb.  A termodinamikai háttér biztosítja robosztus --- azaz a
paraméterek, kezdeti és peremfeltételek váltpzására nem érzékeny --- és stabil numerikus 
eljárások létezését. A képlékenységen túlmenően is a hiszterézis jelenségének megfelelő leírását adja.

Egy, az előzőektől lényegesen különböző, de termodinamikailag következetes
 megközelítést ad \nev{Asszonyi}, \nev{Ván} és \nev{Szarka} \cite{AssAta07b}. 
 Ők a képlékenységi feltételt az entrópiafüggvény változójának
tekintik és ennek segítségével adják meg a rugalmassági állandók képlékenység
következtében bekövetkező változását. Alapfeltevésük, hogy a képlékenységi feltétel
munka alapú. Ez a megközelítés alkalmas  viszkoelasztoplasztikus, képlékeny-reológiai 
növekményes
egyenletek közvetlen le\-vezetésére és a deformáció időbeli változásának leírására reológiai
és képlékeny hatások esetén. A képlékenységi feltétel közvetlen használata
viszont azzal jár, hogy a rugalmassági modulus  a képlékenységi határ átlépésekor
ugrásszerűen változik, lényegében úgy, ahogy  \(C_{ijkl}\) és \re{idplrugmod}
különbözik Az időbeli változások differenciálegyenleteken
alapuló leírásakor vi\-szont ez a deformáció ugrásszerű változását okozza,
ami fizikailag irreális, és ezért további feltételekkel kell kiküszöbölnünk.
Az elmélet kiterjeszthetőségét
nem-kapcsolt  képlékenység leírására nem vizsgáltuk. 

A kapcsolt képlékenység azonban nem jó modellje a talajoknak és köveknek (dilatancia),
nem ad leírást a szöglethatásra, és nem magyarázható vele a lágyulás (softening)
jelensége \cite{JirBaz02b,Mic09b}. A három probléma közül az első kérdéskör,
azaz a
dilatancia leírásával kapcsolatos nehézségek köre  tűnik a legalapvetőbbnek. A
dilatancia azt jelenti, hogy nyírási képlékeny alakváltozás térfogatnövekedést
okoz - mindennapos jelenség sűrű szemcsés anyagokban. Amint
majd látni fogjuk, a kapcsoltság (asszociativitás) alapvető termodinamikai
feltételekhez köthető, így feloldása
feltehetőleg szükségessé teszi a klasszikus termodinamikai képlékenység általánosítását. 
Képlékenység-mechanikai  oldalról ---
ha nem vesszük figyelembe a második főtételt ---, akkor könnyű dolgunk van:
már említettük, hogy a  folyásfügvény és a képlékeny potenciál megkülönböztetése
vezet eredményre.
 Ha viszont legalább nagyjából meg akarjuk tartani a  termodinamikai 
kereteket csak néhány általánosítást ismerünk.

  \nev{Ristinmaa} és \nev{Ottosen} \cite{RisOtt98a} két részre osztják  a belső változókhoz 
konjugált termodinamikai áramokat, és mindkét halmazhoz disszipációs potenciált 
feltételeznek, illetve gyártanak. A többféle potenciállal valójában megsértik a termodinamikai
kereteket. Viszont így levezethetőek szöglethatások (amikor a folyási függvény
szintfelülete törik) és nem-kapcsolt kép\-lékenység
is, többek között \nev{Duvout-Lions} jellegű.
A dinamikai folyásfelületre vonatko\-zó javaslatuk  viszont visszaállítja a
klasszikus képlékenység lazább - és a a második főtétellel tisztázatlan viszonyú
- önkényes feltevéseit \cite{RisOtt00a}, ezzel gyakorlatilag kiküszöböli
a termodinamikai megközelítés előnyeit. 
 
\nev{Houlsby} (és talán \nev{Collins}) hiperképlékenységnek nevezett elméletét a belső változókra
vonatkozó kinematikai kényszerek teszik nem-kapcsolttá \cite{HouPuz06b}. A
továbbiakban kifejtett --- tisztán termodinamikai --- megfontolásokhoz (és a \nev{Ziegler}-féle
elgondolások lényegéhez) ez áll legközelebb.

\section{A súrlódás termodinamikája}

\begin{figure}
\includegraphics[width=0.8\textwidth]{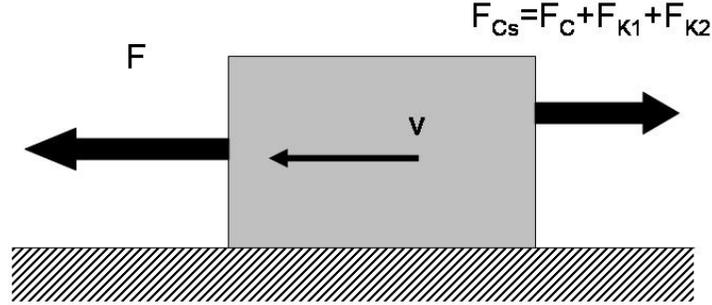}
\caption{\label{fsurl}
Csúszó, súrlódó csillapított mozgású mechanikai-termodinamikai test.
Valójában mechankailag tömegpont, de a pontmechanikai hagyományoknak megfelelően merev testként szemléltettük.  }
\end{figure}

Tekintsünk egy vízszintes talajon \(F\) külső erővel mozgatott \(m\) tömegű
testet, amelyre súrlódási és csillapító erők hatnak (\ref{fsurl}. ábra). 
Az elemi fizikából
jól ismert módon háromféle közegerőt szoktunk feltételezni a mozgás 
fékezőjeként. 
\begin{itemize}
\item Az sebesség nagyságától független Coulomb-féle súrlódási erőt
        \begin{equation} |F_{C} |=\alpha = \mu N =\text{\'all.,}  \end{equation}
\item a sebességgel arányos közegellenállást\ 
        \begin{equation} |F_{K1}| = \beta v,  \end{equation}
\item és a sebesség négyzetével arányos közegellenállást 
        \begin{equation} |F_{K2}| = \gamma v^2  .\end{equation}
\end{itemize}

Ezek mind a mozgást fékező erők, ezért irányuk a sebességgel ellentétes.
Vagyis egy dimenzióban a teljes fékező erő pontosabb formája:
\begin{equation}
        F_{Cs} = F_C+ F_{K1} + F_{K2} =- \alpha \frac{v}{|v|} - \beta v-
        \gamma v |v|.          
\end{equation}

A fenti mozgásra vonatkozó differenciálegyenlet pedig a következő
lesz:
\begin{equation}
       m\ddot x  = F+F_{Cs}= F-\alpha \frac{\dot x}{|\dot x|} - \beta \dot x-
        \gamma \dot x |\dot x|.          
\label{mechlen}\end{equation}

A csillapító erők azonban nemcsak a mozgást fékezik, hanem egyúttal disszipatívak.
Munkájuk melegíti a testet és környezetét. Ez utóbbi tulajdonság termodinamikai
keretek között  válik érthetővé és egyúttal levezethető mozgást fékező természetük. Tehát
vegyük észre, hogy az \re{fsurl} ábrán egy termodinamikai rendszert látunk, egy
környezetével kapcsolatban álló homogén termodinamikai testet. 
  
A termodinamikai tárgyalásban a konstitutív mennyiségekre vonatkozó meg\-szorításokat
kaphatjuk meg. Ennek első lépéseként az alapmérlegeket írjuk fel. Az impulzusmérleg
esetünkben az előbbiekben felírt \re{mechlen} egyenlet, de benne a rendszer
és környezet kapcsolatát leíró, csillapító
erő konstitutív mennyiség, amit a második főtétel követelményeinek megfelelően
szeretnénk előírni:
\begin{equation}
       m\ddot x  = F+F_{Cs}.          
\label{ibal}\end{equation}
Tegyük fel továbbá, hogy a tömegpont \(E\)
energiáját csak a külső \(F\) erő munkája változtatja, annak hiányában megmaradna. Tehát
\begin{equation}
\dot E= F \dot x.
\label{ebal}\end{equation}  Látni, fogjuk, hogy ebből a feltevésből --- termodinamikai
keretek között --- következik, hogy csillapító erők
 munkája
csak a tömegpont belső energiáját növeli. Gondolatmenetünkben az egyszerűség
kedvéért nem foglalkozunk a környezet energiamérlegével. A tömegpont \(U\) belső 
energiája definíció szerint a \(E\) energiájának és a kinetikus és
potenciális energiájának különbsége, azaz
\begin{displaymath}
U= E- m\frac{\dot x^{2}}{2}  .
\end{displaymath} 

Az entrópia a belső energia függvénye, belső energia szerinti deriváltja
pedig a hőmérséklet reciproka. Éppen ezért a \re{ebal} energiamérleg 
és a \re{ibal} impulzusmérleg felhasználásával kapjuk, hogy 
 \begin{equation}
\dot S(U)=\frac{1}{T} \dot U =
     \frac{1}{T}\left(\dot E - m \dot x\ddot x \right)=
     - \frac{\dot x}{T}\left(m \ddot x -F \right) =
     - \frac{1}{T} F_{Cs} \dot x.
\end{equation}

A második főtétel értelmében pedig az entrópia növekszik,  
 \begin{equation}
\dot S = - \frac{1}{T} F_{Cs} \dot x \geq 0,
\label{eegyn}\end{equation}
tehát a csillapító erő ellentétes irányú kell legyen a sebességgel. Ez a
formula kínálja a termodinamikai erők és áramok azonosítását is. A sebességet
a mozgás meghatározza. A csillapító erő viszont a test és a környezet viszonyát megadó függvény, 
tehát ez lesz a meghatározandó termodinamikai áram. Az egyenlőtlenség értelmében, a
csillapító erő, mint
konstitutív mennyiség csak
 a  sebesség
függvénye lehet, ami  pedig jelen esetben független változóként termodinamikai erő. A termodinamikai
és mechanikai elnevezések erre a furcsaságára, vagyis hogy a csillapító
erő termodinamikai szempontból áramnak tekintendő, a kontinuummechanikai tárgyalásban
már felhívtuk a figyelmet \cite{VanAss06a}. A kontinuumok esetén sebességgradiens bizonyult termodinamikai erőnek, itt pedig a sebesség  
(illetve a sebesség és a hőmérséklet hányadosa).
Felhívjuk a figyelmet, hogy a sebesség itt a súrlódó felületek közötti relatív
sebesség és a fenti tárgyalás - megfelelően általánosítva - túllép a termosztatikán
és a mozgó testek közönséges termodinamikájához tartozik. A mozgó homogén
testek termodinamikája a mechanika és a termodinamika egyesítétésének fontos
eleme, számos régi és új megoldatlan és megoldatlannak hitt problémával 
\cite{BirVan10a,Vel10m}. 

 Azonnal láthatjuk, hogy a kontinuumok esetén szokásos szigorúan lineáris erő-áram kapcsolat
csak egyik, a sebességgel arányos,  fajtáját adja a jól ismert csillapítási erőknek. A \nev{Coulomb}-
súrlódás és a sebesség négyzetével arányos csillapítás szintén megfelel a
\re{eegyn}
egyenlőtlenségnek, a következő módon:
\begin{equation}
 F_{Cs} = -L(\dot x) \dot x  = -\left(\beta + \alpha \frac{1}{|\dot x|} +
 \gamma |\dot x|\right)\dot x
\label{ons1}\end{equation}

Az erő-áram kapcsolat nemlineáris, a sebességtől, mint termodinamikai erőtől függ
az \(L\) vezetési együttható. 
A második főtétel egyenlőtlensége megköveteli, hogy az $\alpha, \beta, \gamma$
együtthatók ne legyenek negatívok. Az entrópia növekedésének feltételéből
következtettünk a csillapító erő irányára, utólag igazolva fékező tulajdonságát. 
A klasszikus irreverzibilis termodinamika a \nev{Fourier}-hővezetés, \nev{Fick}-diffúzió, 
vagy például a  \nev{Navier-Stokes}-egyenlet levezetésénél szigorúan lineáris (állandó), vagy 
kvázilineáris (alapváltozóktól függő)
erő-áram kapcsolatot feltételez. Itt most példát láthatunk ettől általánosabb,
nemlineáris vezetési egyenletre, ráadásul egy nagyon egyszerű és jól ismert
jelenségkör esetére. 

\begin{figure}
\includegraphics[width=0.495\textwidth]{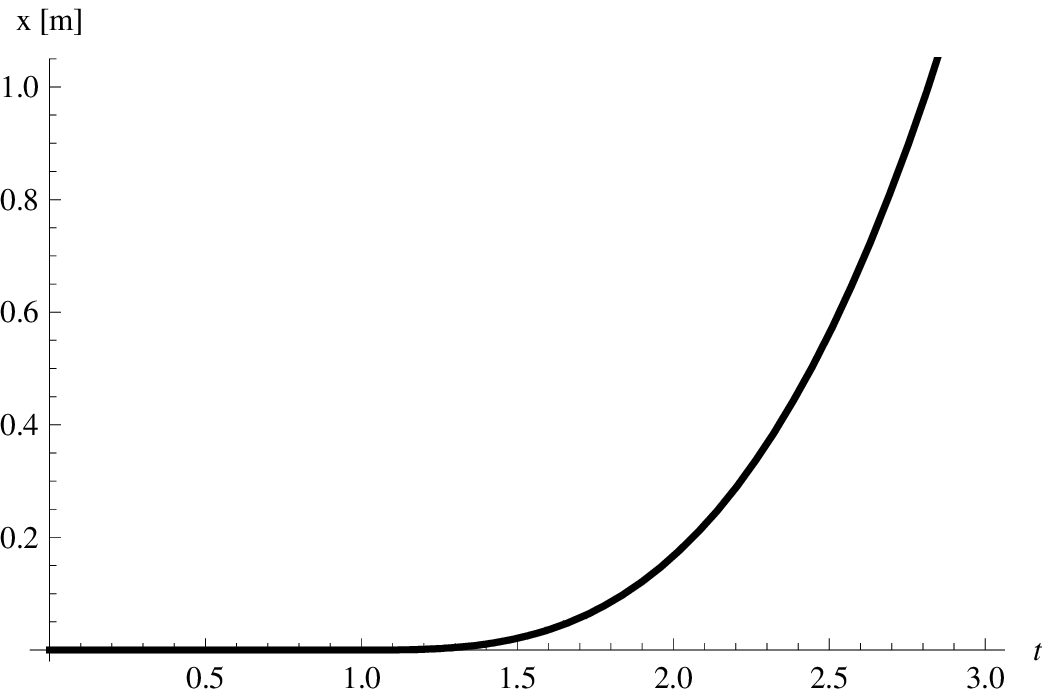}
\includegraphics[width=0.495\textwidth]{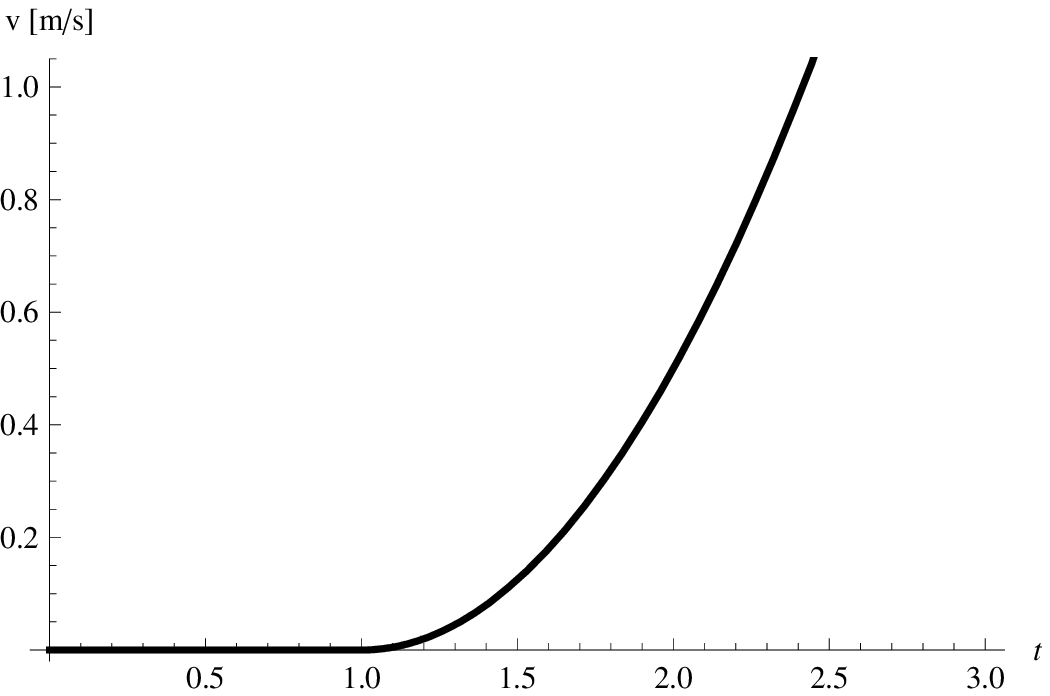}
\caption{\label{nf1}
Az elmozdulás és a sebesség időfüggése a \re{mechlen1} differenciálegyenlet szerint,
$m=1kg$, $V=1m/s$, $\alpha = 1N$, $\beta=0
\frac{kg}{s}$ paraméterértékekkel  számolva. 
A kritikus tapadási súrlódási erő eléréséig a test nem mozdul el és sebessége
is nulla.}
\end{figure}

\begin{figure}
\includegraphics[width=0.8\textwidth]{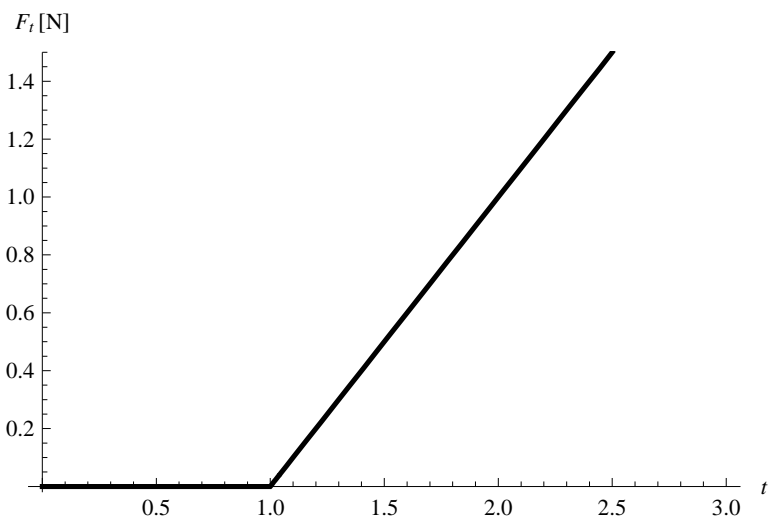}
\caption{\label{nf2}
A teljes erő, azaz a \re{mechlen1} differenciálegyenlet jobb oldalának időfüggése, a  \ref{nf1}. ábra paramétereivel. } 
\end{figure}

A fenti \re{ons1} általános csillapítási anyagtörvény és a belőle
következő  \re{mechlen} differenciálegyenlet azonban fizikailag
nem teljesen felel meg az elvárásainknak és a mindennapi kísérleti tapasztalatnak,
ugyanis nem ad számot a tapadási súrlódásról. Tegyük fel ugyanis, hogy a tömegpontra
 ható erőt egyenletes sebességgel  növeljük, azaz
legyen $F= V t$, ahol \(V\) a terhelési sebesség és \(t\) az idő. Az egyszerűség
kedvéért legyen továbbá $\gamma=0$. Ekkor látjuk, hogy  megoldandó lenne
a
\begin{equation*}
       m\dot v(t)  =V t -\alpha \frac{v(t)}{|v(t)|} - \beta v(t)          
\end{equation*} 
differenciálegyenlet a $v(0)=0$ kezdeti feltétellel.
A matematikai feladat azonban így, a további fizikai feltevések nélkül határozatlan, illetve
fizikailag rossz eredményt ad. Egyrészt a \nev{Coulomb}-súrlódási tag $t=0$-beli
értékét rögzítenünk kell, másrészt fel
kell tételeznünk, hogy a súrlódási erők nem gyorsítják a tömegpontot, azaz
például előírni, hogy a kezdeti feltétel legyen $v(\alpha/V)=0$, a sebesség
növekedése csak a Coulomb súrlódási erő elérése után kezdődik meg. Vagyis
a differenciálegyenlet a következő formában pontosabb fizikai modell:
\begin{equation}
       m\dot v(t)  =
       \begin{cases}
                0, \hskip 3.2cm\text{ha \quad} V t <\alpha,   \\
       V t -\alpha \frac{v(t)}{|v(t)|} - \beta v(t), \quad \text{ha \quad} V t \geq\alpha.
       \end{cases}        
\label{mechlen1}\end{equation} 
Vegyük észre, hogy egy kritikus erő jellegű feltételt adtunk meg, a képlékenység
klasszikus elméletéhez hasonlóan. A differenciálegyenlet megoldásait a 
\ref{nf1}-\ref{nf2}. ábrákon szemléltettük.  

Egy másik, trükkösebb módon viszont a tapadás feltételét eleve tartalmazza
termodinamikai vezetési törvény. A második
főtétel egyenlőtlenségének megoldását ugyanis nemcsak \re{ons1}-hez hasonló,
hatványsorszerű formában kereshetjük. Tegyük fel ugyanis, hogy 
\begin{equation}
 F_{Cs} = -\hat L(F_{Cs}) \dot x  = -(\hat B + \hat A |F_{Cs}|)\dot x,
\label{onsm1}\end{equation}
vagy ezzel lényegében ekvivalensen
\begin{equation}
 \dot x = -L(\dot x)F_{Cs}  = -(B + A |\dot x|)F_{cs}.
\label{onsm2}\end{equation}

Pontosabban, fenntartva a függvénykapcsolatok előre rögzített formáját, azaz
hogy \( F_{Cs}(\dot x)\) határozatlan függvényt keresünk, azt írhatjuk, hogy 
\begin{equation}
 F_{Cs} = - L(\dot x) \dot x  = -\frac{\beta \dot x}{1 + \beta|\dot
 x|/\alpha}.
\label{onsmi}\end{equation}

\begin{figure}
\includegraphics[width=0.8\textwidth]{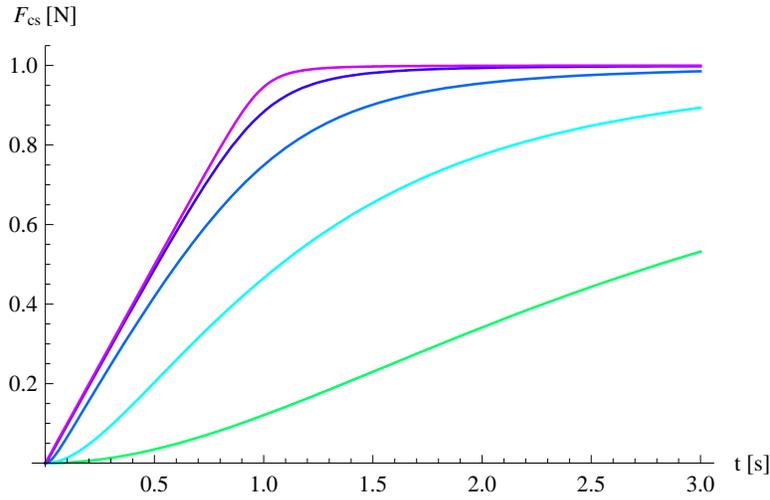}
\caption{\label{f1}
A csillapítási erő időfüggése. $m=1kg$, $V=1m/s$, $\alpha = 1N$, $\beta=\{0.3, 3, 30, 300, 3000\}
\frac{kg}{s}$. A legfelső, legszögletesebb görbéhez tartozik a legnagyobb
$\beta=3000\frac{kg}{s}$ érték. Nagy $\beta$ esetén a súrlódási erő a terhelőerővel
együtt növekszik, majd állandó értéket vesz fel. } 
\end{figure}

\begin{figure}
\includegraphics[width=0.495\textwidth]{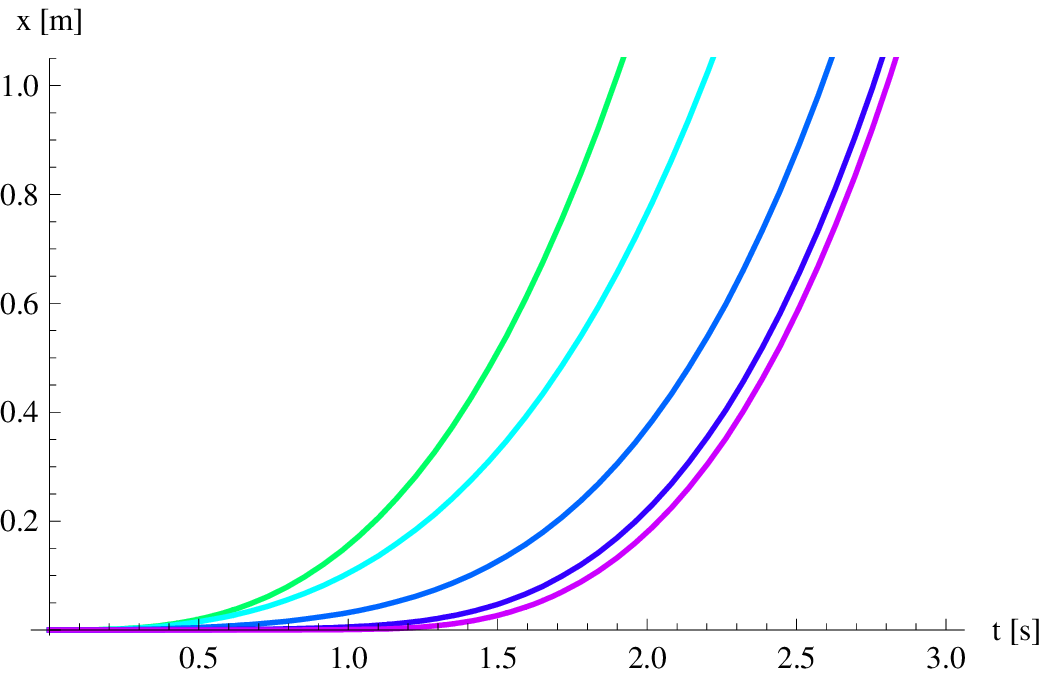}
\includegraphics[width=0.495\textwidth]{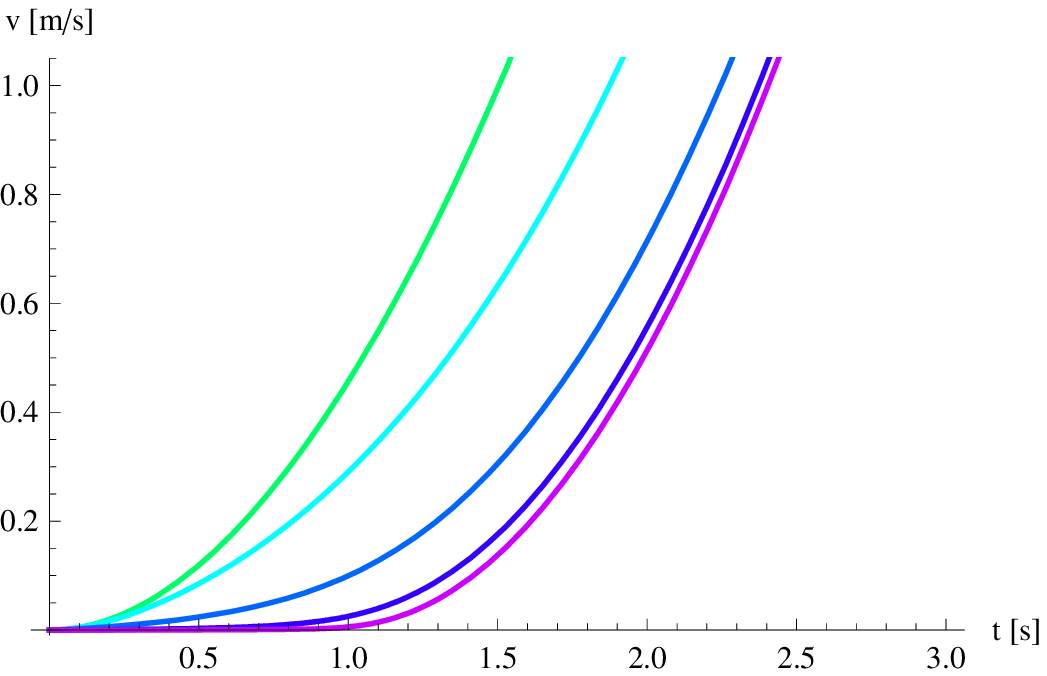}
\caption{\label{f2}
Az elmozdulás és a sebesség időfüggése az 1. ábra paramétereivel  számolva. 
A legfelső,  görbéhez tartozik a legkisebb $\beta=0.3\frac{kg}{s} $ érték.
Nagy $\beta$ esetén a test nem mozdul el és sebessége nagyon
kicsi amíg a terhelés el nem éri az $F=\alpha$  határt.}
\end{figure}

A fenti formulákban $\hat A, \hat B, A, B$, $\alpha=1/A$ és  $\beta=1/B$ anyagi  
paraméterek. A második főtétel egyenlőtlenségét anyagi tulajdonságként, 
tehát a folyamatoktól (azaz jelen esetben $\dot x$-től) függetlenül megkövetelve, 
egyik fenti paraméter sem lehet negatív. Ha $1\gg\beta|\dot x|/\alpha$, akkor  
$F_{cs} \approx\ -{\beta \dot x}$. Ha $1\ll\beta|\dot x|/\alpha$, akkor  
$F_{cs} \approx\ -\alpha \dot x/|\dot x|$. Tehát kis sebességek esetén a
sebességgel
arányos a súrlódási erő, nagyobb sebességek esetén állandó. Egyenletesen növekvő erő hatásának
kitett súrlódással
mozgó tömegpont mozgásegyenlete tehát
\begin{equation}
       m\dot v(t)  =V t - \frac{\beta v(t)}{1+ \frac{\beta}{\alpha}|v(t)|}.          
\label{mechlen2}\end{equation} 
Ez az együtthatók és az időskála megfelelő beállításával pontosan az elvárt tapadási-csúszási
súrlódásos viselkedést eredményezi (\ref{f1}-\ref{f2} ábrák). Az előző, klassszikus súrlódásra
vonatkozó \re{mechlen1} differenciálegyenlethez képest \re{mechlen2} láthatóan
számot ad a mozgás kezdetéről, csak $\beta\rightarrow \infty$ esetén kapjuk
vissza az előző egyenlet megoldásait, illetve \re{mechlen1} éles feltételét.  

Másrészt viszont ami az előnye ennek a modellnek, az egyúttal a hátránya
is. Ugyan az éles átmenet helyett egy, a $\beta$ paraméterrel hangolható, tompított 
álló-csúszó (merev-képlékeny) átmenetet kapunk, de a test már az erőhatás
kezdetétől fogva mozog egy kicsit. A fizikai képünk is ennek megfelelően
változik: ez  a fajta  sebességgel arányos csillapítás kis sebességek esetén
érvényes.
 
A súrlódás jelenségének fenti nagyon egyszerű modelljét szemléltetésnek
szántuk. Nem vettük figyelembe az erő irányát és a nyomóerő hatását sem.
A jelenségkörnek azonban ezeken felül is számos további olyan vonatkozása 
van, amelynek nincs megfelelő termodinamikai leírása \cite{RuiPra02b,PerVol02a}. 
A súrlódási konstitutív törvények
termodinamikai megszorításainak vizsgálata önmagában is érdekes, mert elkülöníti
a jelenségkör univerzális és anyagfüggő vonatkozásait. Az első lépéseket
ebben az irányban \nev{Verhás} írása jelenti \cite{Ver09m}. Valódi anyagi paraméterek
azonosítása a kontaktmechanikában (a súrlódási, gördülési és ütközési tulajdonságok
vizsgálatakor) önmagában is fontos, ráadásul a képlékenységi 
elméletek mélyebb megértéséhez és továbbfejlesztéséhez vezethet. \re{onsm2}-\re{onsmi} 
vezetési egyenletekhez tartozó erő-áram kép, azaz az egyenletek felállítása
és ennek megfelelően az általánosításának iránya is különböző, de a kapott mozgásegyenletek
ekvivalensek. \re{onsm1} nem ekvivalens az
utóbbiakkal, de a képlékenység irodalmában mégis elsősorban ez  a forma
terjedt el.  Az utolsó fejezetben látni fogjuk, hogy a tapadási viselkedést
tükrözi, nagyon hasonló megoldásokra vezet, mint \re{onsm2}-\re{onsmi}.

\section{A reológia termodinamikai elmélete - kis deformációk}

\subsection{Mérlegek} A klasszikus  képlékenységelméletek termodinamikai megalapozottságának
hiánya  különösen
a reológiával történő összevetés fényében szembetűnő. A termodinamikai
reológia egyenletei és egész anyagelmélete - azaz a konstitutív egyenletek származtatási módja
- ugyanis világos módon a második főtételen alapul.

 A reológia gyakorlatban leginkább
használt alapmodelljei az empirikus alapon származtatott skalár, lineáris
elemek ad hoc kapcsolásából adódnak. Az ilyen 'félempirikus'
modellezés
teljesítőké\-pessége azonban korlátozott, ugyanis általában nem anyagmodellekről van szó bennük,
hanem inkább körülménymodellekről, mert paramétereik függenek a körülményektől 
(pl. terhelési feltételektől,  irányoktól és sebességektől). A
valódi anyagi paraméterek és mo\-dellek keresése vezetett az objektivitást
(vonatkoztatási rendszertől való függetlenséget) és a termodinamikai követelményeket érvényesítő elméletek kidolgozásához.
Hiába egyszerűek a klasszikus reológia  skalár lineáris  egyenletei,
ha érvényességük korlátozott volta miatt a megfelelő  paramétereket mindig újra és újra (esetleg speciális gépekkel)
meg kell mérnünk. Van, amikor ez lehetetlen vagy
költségesebb, mint egy megfelelően megbízható, kevésbé körülményfüggő anyagmodell használata.
Ezért az általános elvi követelemények\-nek is megfelelő, éppen ezért sokkal
szigorúbb keretekben kidolgozott modellek iránti igény nem csak esztétikai,
hanem végső soron gyakorlati, gazdaságossági követelmény.

A  reológia két 
alapjelensége a kúszás és a relaxáció. Mechanikai testet ugrás\-szerűen megterhelve
és a terhelést ezután állandóan tartva a deformáció ugrás\-szerű kezdeti változás
után fokozatosan veszi fel állandósult értékét: ez a kúszás. Mechanikai testet ugrás\-szerűen deformálva
és a deformációt állandó értéken tartva a feszültség fokozatosan veszi fel
állandósult értékét:\ ez a (feszültség)relaxáció. Mindkét jelenség magyarázata
a rugalmasságtanon túlmutat és  viszkoelaszticitás, hipoelaszticitás
nevek alatt találhatóak meg  őket a mechanikai irodalomban.
A nehézséget általában a két alapjelenség egy modell keretein belül történő,
egységes értelmezése jelenti.

Az elvi, többek között termodinamikai követelményeknek megfelelő első elmélet,
 a már említett,\ \nev{Truesdell} és \nev{Noll} nevéhez köthető, hiporugalmasság, amely feltételezi, hogy
a fe\-szültségtenzor nemcsak a deformáció, hanem a deformáció időderiváltjának
is függ\-vénye. Azonban a termodinamikai potenciálok létezése nem dobható
el következmé\-nyek nélkül, a hiporugalmasság elmélete túlságosan laza, ezért könnyen
vezet rossz anyagfüggvényekre.

Az első, termodinamikailag igazán megfelelő, nagy deformációkra vonatkozó és objektív időderiváltakat
használó, belső változókon alapuló reológiai elméletet \nev{Verhás} dolgozta ki
\cite{Ver97b}, \nev{Kluitenberg} \cite{Klu77a,KluCia78a} úttörő munkáira alapozva. E 
szerint az elmélet szerint
a relaxáció és a kúszás egyenrangú és az ezeket egyszerre tartalmazó alapmodellt,
a \nev{Poynting-Thomson}-féle ún. standard modellt egyetlen belső változó segítségével 
megkaphatjuk termodinamikailag minimálisnak tekinthető további feltételekkel.

Az alábbiakban röviden összefoglaljuk \nev{Verhás} elméletét, kis deformációk esetére
szorítkozva. Az
elméletnek többféle nagy deformációs kiterjesztése is létezik, amelyek az
objektivitás követelményét is figyelembe veszik.  Ezeknek az elvi szempontból teljesnek tekinthető elképzeléseknek azonban
a fent említett
egyszerű reológiai alapjelenségeken túlmutató, kísérletekkel
történő összevetése máig nem teljes; több probléma megoldásra vár. Megjegyezzük,
hogy a reológiának jelenleg\textit{ nincs }olyan modellje - sem olyan, amely
megfelel a fenti két elvi követelménynek (az objektivitásnak és a termodinamikai
következetességnek), sem másmilyen -, amely minden fő reológiai kísérlet
(egyszerű nyírás, viszkozitás, nyírási relaxáció, nyírási szünet, stb...) során
kielégítő egyezést mutatna a mérésekkel. \
  
Minden képlékeny és reológiai modell felállításakor az alapmérlegek felírásából  és 
entrópiaprodukció levezetéséből indulunk ki. Esetünkben az entrópiaprodukció 
kiszámításához a  tömeg-, lendület- és energiamérlegeket kell figyelembe
vennünk. 

A tömeget megmaradónak tekintve kapjuk, hogy
\begin{equation}
  \dot\rho +\rho \partial_i v^i = 0,
\end{equation}  
ahol $\rho$ a sűrűség, $v^i$ a sebességmező, a pont pedig a szubsztanciális
időderiváltat jelöli. Az előzőekhez hasonlóan
indexes írásmódot alkalmazunk. A lendületmérleg formája pedig a következő
lesz, ha eltekintünk a külső, térfogati erőktől, amelyek nem játszhatnak
szerepet az anyagegyenletek levezetésénél:
\begin{equation}
  \rho\dot v^i - \partial_j t^{ij} = 0^i.
\end{equation}

Itt $t^{ij}$ a feszültségtenzor. Feltételezzük, hogy a közegben nincs belső impulzusmomentum, 
tehát a feszültségtenzor szimmetrikus: $t^{ij}=t^{ji}$. A teljes energia mérlege
 \begin{equation}
  \rho\dot e  + \partial_i q_{t}^i = 0,
\end{equation}
ahol $e$ a teljes energia fajlagos értéke, $q_{t}^i$ pedig az áramsűrűsége. Egykomponensű 
egy\-szerű mechanikai kontinuumok - elsősorban folyadékok - esetén a 
belső energia a teljes és a kinetikus energia különbsége. Ezt a definíciót
használtuk az előző fejezet súrlódásra vonatkzó megfontolásaiban is.   A rugalmasságtanban 
a mechanikai energiát általában a (\nev{Helmoltz}-féle) szabadenergia segítségével
kötik a termodinamikai követelményekhez. Reológiai rendszerekben feltételezik, hogy a 
mechanikai hatást az anyagban
végbemenő strukturális változások késleltetik. Ezt  a memória-, illetve tehetet\-len\-ségi
jelenséget egyetlen szim\-metrikus másodrendű tenzor dinamikai változóval veszik figyelembe.
Most a belső e\-nergiát a teljes energia és a más energiafajták (kinetikus,
rugalmas) különbségeként fogjuk értelmezni. Mint látni fogjuk, ez egy\-szerűsítéseket
jelent a tárgyalásmódban és könnyen megmutatható, hogy izoterm esetben ekvivalens
a hagyományos, \nev{Verhás} által is alkalmazott megoldással, ahol az entrópiát
egészítik ki a dinamikai (vagy belső) változók kvadratikus formájával \cite{Ver85b}.   
Az ideálisan rugalmas izotrop kontinuum fajlagos rugalmas energiája
kis deformációk esetén\begin{equation}
e_{rug} = \frac{\lambda}{2} (\ep^{i}_{\;i})^2 + \mu \tilde \epsilon^{ij}\tilde\epsilon_{ij},
\label{erug} \end{equation}
ahol $\lambda$ és $\mu$ a \nev{Lamé}-állandók, $\tilde \ep^{ij} = \ep^{ij} - \ep^{k}_{\;k}/3
\delta^{ij}$ pedig a deformáció nulla nyomú része ($\ep^{i}_{\;i}$ a nyoma indexes
jelöléssel).

A dinamikai változót \(\xi^{ij}\)-vel jelöljük és hatását figyelembe vesszük
a teljes energia meghatározásakor. Feltételezzük, hogy járuléka az eddigi
energiafélékhez hasonlóan additív és független, formája pedig a kinetikus
energiáéhoz hasonlóan - nem véletlenül - kvadratikus. Ebből következően
az $e_B$ belső e\-nergia
\begin{equation}
  e_B= e- \frac{v^2}{2}- e_{rug}(\ep^{{ij}}) -e_{din}(\xi^{{ij}}) .
\label{rben}\end{equation}
 
Az első két tagot, a teljes és a kinetikus energia különbségeként 
meghatározott szokásos belső energiát \(e_b\)-vel jelöljük: 
\begin{equation}
  e_b= e- \frac{v^2}{2}.
\end{equation}

A belső, dinamikai változókat nem a belső energiát, hanem csak az entrópiát 
módosító módon szokás figyelembe venni (lásd pl. \cite{Ver85b,JouAta92b}). 
A kétféle megközelítés
egyenértékű, illetve a hőmérséklet szerepét illetően az energiát módosító javaslat 
fizikailag világosabb. Ez
a belső változóhoz köthető kiegészítő energia jelzi, hogy a változó segítségével
modellezett hatásnak tehetetlensége van. Ha a tehetetlenségi hatások mikroszkopikus,
strukturális mechanizmusát
is ismernénk, akkor célszerű lenne bevezetni a tehetetlenséget jellemző,
belső változóhoz kötődő tömegszerű együtt\-hatót. Ennek hiányában a \nev{Morse}-lemma 
értelmében  izotróp anyagra a belső változóval reprezentált anyagi
kinetikus energia járulék általánosan tiszta négyzetes formában írható, mert a
skálát nincs okunk bármihez kötni, azaz:

\begin{equation}
e_{din}(\xi^{{ij}}) =\frac{1}{2} \xi^{ij}\xi_{ij}. 
\label{edin}\end{equation}  
A hagyományos \(e_{b}\) belső energia fluxusára, azaz konduktív áramsűrűségére vo\-natkozóan
pedig a következő (szokásos) összefüggést feltételezzük
\begin{equation}
q^i = q_{t}^i - t^{ij} v_j.
\end{equation}
Ekkor az \(e_b\) belső energia mérlege 
\begin{equation}
  \rho\dot e_b + \partial_j  q^j =  t^{ij}\partial_jv_i.
\end{equation}

\subsection{Entrópia és mérlege} 
Az entrópiamérleg felírásakor az entrópiasűrűség változóinak  megállapítása,
illetve a konduktív entrópia-áramsűrűség formájának megtalálása az alapvető feladat. 
Klasszikusan, gázok és folyadékok esetén
ezek az extenzív változók sűrűségei. A kontinummechanikában a belső energia
sűrűsége és valamilyen objektív
deformációmérték a leggyakoribb választási lehetőség. Kis deformációs közelítésben
ezek egyenértékűek. Az előbbiekben már kiválasztottuk az entrópia változóit,
és mindet a belső energián keresztül vettük figyelembe,
azaz   \(s(e_{b},\epsilon^{ij},\xi^{ij})=\hat s(e_B)\). Ennek megfelelően \re{erug}--\re{edin}
felhasználásával a \nev{Gibbs}-reláció  a következő:
\begin{equation}
 de_B  = de_b-({\lambda}\ep^{k}_{\;k}\delta^{ij}+ \mu \tilde \epsilon^{ij})d\epsilon_{ij}-
 \xi^{ij} d\xi_{ij}= Tds+ (t_{s})^{ij} d\epsilon^{ij}-
 \xi^{ij} d\xi_{ij}.
\end{equation}  

Itt $(t_s)^{ij}=\pd{e_{rug}}{\ep_{ij}}={\lambda}\ep^{k}_{\;k}\delta^{ij}+2 \mu 
\tilde \epsilon^{ij}$ a termosztatikai feszültségtenzor, $T$ a hőmérséklet. Innét 
leolvasható (illetve tulajdonképpen az intenzív
mennyiségeket és rajtuk keresztül az entrópiát a parciális deriváltjain
keresztül definiálja), hogy 
$$
\pd{\hat s}{e_B} =\pd{s}{e_b}= \frac{1}{T}, \qquad 
\pd{s}{\epsilon^{ij}} =-  \frac{1}{T}\pd{e_{rug}}{\epsilon^{ij}}=-\frac{(t_s)^{ij}}{ T},
\qquad
\pd{s}{\xi^{ij}} =-  \frac{1}{T}\pd{e_{din}}{\xi^{ij}} =
        -  \frac{\xi_{ij}}{T}.
$$

Ezek a parciális deriváltak a termosztatikai intenzív mennyiségeknek felelnek
meg. A belső változónk a szokásos értelemben nem biztos, hogy extenzív (általában
ennek a tulajdonságnak nincs nagy jelentősége belső változókra), de a hozzá
tartozó entrópiaderivált nulla volta definiálja a termodinamikai egyensúlyt.
Mivel ennek értéke a fenti utolsó formulából láthatóan $\xi_{ij}$-vel arányos,
ezért a belső változó egyúttal \textit{\nev{Verhás}-féle dinamikai szabadsági fok }is, mert
termodinamikai egyensúlyban értéke nulla. Dinamikai szabadsági fokok lehetnek
például a
kiterjesztett termodinamikában független változóként bevezetett termodinamikai
áramok \cite{Gya77a,JouAta92b}, vagy bizonyos értelemben a relatív impulzus
is \cite{VanAta08a}. 

Az entrópia konduktív áramsűrűsége
a klasszikus választás szerint a hőáramsűrűség és a hőmérséklet hányadosa: \(j^i = q^i/T\).
Míg az entrópia változóinak kiválasztása a fizikai modellezés része, az entrópia
áramának formája a modern kontinuum-termodinamika módszereivel kiszámolható.
Itt most ezt a levezetést nem adjuk meg, mert egy hosszasabb matematikai
módszer (pl. \nev{Liu-} vagy a \nev{Coleman-Noll}-eljárás) alkalmazásával 
csak a jól ismert szokásos eredményre
jutnánk. Általában bizonyítható, hogy lokálisan egyensúlyi, elsőrendűen 
gyengén nemlokális irreverzilis termodinamikában
--- azaz a mi esetünkben is --- egykomponensű közegekre az alapmérlegek 
és az entrópiamérleg egyenlőtlenségének következményeként az
entrópia-áramsűrűség általában
a hőáramsűrűség és a hőmérséklet hányadosa\ \cite{Van03a}.

Az entrópiaprodukció ezek után a következő:
\begin{gather}
  \rho\dot s+ \partial_j  (j_s)^j =  
  \rho\dot s(e_{b},\epsilon^{ij},\xi^{ij})+ \partial_j  \frac{q^j}{T} =\nonumber\\
  -\frac{1}{T}(\partial_j q^j-t^{ij}\partial_jv_i)- 
        \frac{(t_s)^{ij}}{T}  \dot\epsilon_{ij}-  \frac{\xi_{ij}}{T}\dot \xi_{ij}+
        \partial_j  \frac{q^j}{T}= \nonumber\\ 
   \frac{1}{T}\left(t^{ij}-(t_s)^{ij}\right)\dot\epsilon_{ij}-  \frac{\xi_{ij}}{T}\dot \xi_{ij}+ 
   q^i \partial_i \frac{1}{T} \geq 0.
\end{gather}

Itt felhasználtuk a kis deformációk esetén a sebességgradiensre érvényes
összefüggést:
\begin{equation}
\partial_i v_j = \dot \epsilon_{ij}.
\end{equation}
 
Ezek után a mechanikai folyamatokra szorítkozva feltételezzük, hogy a hőmérsékletel\-oszlás
homogén a kontinuumban, vagy a hőáramsűrűség nulla (izoterm, illetve adiabatikus
folyamatok). Ekkor a fenti formula utolsó tagja nulla és az  energiadisszipáció,
azaz az entrópiaprodukció szorozva a hőmérséklettel a következő:
\begin{equation}
 T\sigma_{s}= \left(t^{ij}-(t_s)^{ij}\right)\dot\epsilon_{ij}-{\xi_{ij}}\dot \xi^{ij}
    \geqslant 0.
\end{equation} 

\subsection{Vezetési (konstitutív) egyenletek} A klasszikus irreverzibilis
termodinamikában az entrópiaprodukció
segítségével termodinami\-kai erőket és áramokat
azonosítunk, és közöttük lineáris kapcsolatot feltétele\-zünk. Ezzel megoldjuk az egyenlőtlenséget. Ahogy
már az előző fejezetben is említettük, az áramok és erők megkülönböztetése nem önkényes és nincs benne fizikai feltevés, inkább matematikai jellegű. A termodinamikai 
erők az állapot ismert függvényei, 
az áramok pedig konstitutív - azaz meghatározandó - függvényeket tartalmaznak.
Jelen esetben a feszültség és a belső változó evolúciós egyenlete a határozatlan,
tehát
\begin{center}
\begin{tabular}{c|c|c}
Erő & $\dot\ep^{ij}$ & $-\xi^{ij}$ \\\hline
Áram & $t^{ij}-(t_s)^{ij}$ & $\dot\xi^{ij}$ \\
\end{tabular}
\end{center}

A lineáris vezetési egyenletek ezek után a következők lesznek:
\begin{eqnarray}
t^{ij}-(t_s)^{ij} &=& L_{11}^{ijkl} \dot\ep_{kl}-  L_{12}^{ijkl} \xi_{kl},\label{o1}\\
\dot\xi^{ij} &=&  L_{21}^{ijkl}  \dot\ep_{kl} -  L_{22}^{ijkl} \xi_{kl}.
\label{o2}\end{eqnarray}

Itt \(L_{11}\), \(L_{12}\), \(L_{21}\) és \(L_{22}\) negyedrendű csatolási
mátrixok, amik izotrop esetben 2-2 skalár együtthatót tartalmaznak a szimmetrikus
tenzor deformáció és belső változó gömbi és deviatorikus részének megfelelően.
Ekkor a fenti egyenletrendszer is két független részre esik szét:
\begin{eqnarray}
t^i_{\;i}-(t_s)^i_{\;i} &=& 
        m_{11} \dot\ep^i_{\;i} - m_{12} \xi^i_{\;i},\label{io1}\\
\dot\xi^i_{\;i} &=&  
        m_{21} \dot\ep^i_{\;i}  -  m_{22} \xi^i_{\;i},\label{io2}\\
\tilde t^{ij}-(\tilde t_s)^{ij} &=& 
        k_{11}\dot{\tilde  \ep}^{ij}-  k_{12}\tilde \xi^{ij},\label{io3}\\
\dot{\tilde  \xi}^{ij} &=&  
        k_{21}\dot{\tilde  \ep}^{ij} -  k_{22}{\tilde  \xi}^{ij}.
\label{io4}\end{eqnarray}

 A belső változók általában kiküszöbölhetők a fenti \re{o1}--\re{o2} illetve
a  \re{io1}--\re{io2} egyenlet\-rendszerekből. Külön-külön a deviatorikus
és a térfogati részekre egy-egy ún. tehetetlenségi \nev{Poynting-Thomson}-modellt
eredményeznek \cite{Ver85b,VanAss06a}. 

Ennek megfelelően az entrópiaprodukció is kvadratikus lesz a termodinamikai erőkben vagy áramokban,
 és a skalár és másodrendű szimmetrikus nyomnélküli tenzorok
szét\-csatolódnak:
\begin{multline}
 T\sigma_{s} = 
        m_{11} (\dot\ep^{i}_{\;i})^2 -  
        (m_{12}+m_{21})\dot\ep^{i}_{\;i}\xi^{j}_{\;j} +
        m_{22} (\dot\xi^{i}_{\;i})^2 + \\ +
        k_{11} \dot{\tilde  \ep}^{ij}\dot{\tilde  \ep}_{ij}-
        (k_{12}+k_{21}) \dot{\tilde  \ep}^{ij}\dot{\tilde  \xi}_{ij} +
        k_{22} \dot{\tilde  \xi}^{ij}\dot{\tilde  \xi}_{ij}.
\end{multline} 
 
 A továbbiakban egy térdimenziós tárgyalásra térünk át. 

\subsection{Közönséges reológia - avagy reológia homogén termodinamikai testekre}

Egy térbeli dimenziós eset többféleképpen is adódik a fenti egyenletekből.
Egytengelyű terhelés, vagy csak a térfogatváltozás tárgyalása is egyetlen skalár egyenletre vezet. Ebben az esetben
az eredő egyenlet együtthatói a fenti anyagi paraméterek kombinációi
lesznek. Ha az impulzusmérleget nem vesszük figyelembe, akkor a termodinamikai
egyenletek csak az időbeli változásokra szorítkozva közönséges differenciálegyenletek,
homogén kontinuumra vonatkoznak, ezért ezek az esetek a közönséges, avagy a homogén 
testekre vonatkozó termodinamikának --- a klasszikus termosztatika nemegyensúlyi 
kiterjesztésének --- részét képezik \cite{Mat05b}.     

Összefoglalva a fenti egyenleteket  azt
kapjuk, hogy az energiadisszipáció
\begin{equation}
 T\sigma_{s}= \left(t - \partial_\ep e_r\right)\dot\epsilon-\partial_\xi
 e_r\dot\xi \geqslant 0,
\label{reos}\end{equation} 
ahol eddigi jelöléseinken rövidítve $e_e=e_{rug}+e_{din}$ a rugalmas és a belső változóhoz köt\-hető energia összege: 
\begin{equation}
e_e\ = G \frac{\ep^2}{2} + \bar G\frac{\xi^2}{2}. 
\label{reoer}\end{equation}
Itt $G$ a megfelelő rugalmassági állandó, pl. a \nev{Young}-modulus, ha egytengelyű terhelést tekintünk.  
$\bar G$ a belső változóra vonatkozó
analóg anyagi paraméter. A kontinuumegyenletekben nem vezettük be, mert
csak reológiai jelenségeknél  nincs jelentősége. A képlékenység analóg
tárgyalásakor azonban lényeges lesz, mert ekkor a belső változónak konkrét fizikai jelentése
van: képlékeny deformációként azonosítjuk.
A termodinamikai erők és áramok \re{reos} alapján
\begin{center}
\begin{tabular}{c|c|c}
Erő & $\dot\ep$ & $-\bar G\xi$ \\\hline
Áram & $t^{v}=t-G \ep$ & $\dot\xi$ \\
\end{tabular}
\end{center}

A lineáris vezetési egyenletek ezek után a következőek lesznek:
\begin{eqnarray}
t^{v} &=& l_{1} \dot\ep-  l_{12} \bar G\xi,\label{lo1}\\
\dot\xi &=&  l_{21} \dot\ep-  l_{2} \bar G\xi.
\label{lo2}\end{eqnarray}

A  belső változót kiküszöbölve a fenti (\ref{lo1})-(\ref{lo2})
egyenletrendszerből:
\begin{equation}
\sigma+\tau \dot\sigma= 2\eta\tau_d \ddot \ep+2\eta\dot\ep+2G\ep, 
\label{ptj}\end{equation} 
ahol \(\tau=(\bar Gl_{2})^{-1}\), \(2\eta\tau_d=l_{1}(\bar Gl_{2})^{-1}\),  
\(2 \eta = (l_{1}l_{2}-l_{12}l_{21})l_{2}^{-1}\). Ez az ún. tehetetlenségi 
\nev{Poynting-Thomson}-modell, a minimális modell, amely egyszerre
képes számot adni a relaxációról és a kúszásról is, illetve figyelembe veszi
az anyagi tehetetlenséget. Figyelemre méltó, hogy
egyetlen dinamikai változós, minimális termodinamikai elmélet ezt adja alapmodellként.

Vegyük észre, hogy a fenti modellnek a kereszteffektusok jelenléte, azaz
$l_{12}$ és \(l_{21}\) nem nulla volta lényeges elemét képezi két
kiemelendő szempontból is. Egyrészt ha nincs kereszteffektus, akkor  (\ref{lo1})-(\ref{lo2})
két független egyenlet, a belső változónak nincs hatása a mechanikai jelenségekre,
és nem küszöbölhető ki. Viszont (\ref{ptj})-ben, a kiküszöbölés után,  a kereszeffektusok együtthatóit
már nullának tekinve is érvényes egyenletünk van. Másrészt, $l_{12}$ és \(l_{21}\) viszonyáról
semmit sem feltételeztünk, szándékosan. A szimmetrikus vagy antiszimmetrikus
vezetési mátrixok kérdése egyrészt a teljes termodinamikai képlékenységelméletnek
is egy kulcskérdése (sokak szerint ez dönti el, hogy a képlékenységi modell
kapcsolt-e, vagy nem \cite{Mau92b}), illetve az \re{lo1}-\re{lo2} vezetési
egyenletekben \(\bar G\) értékének megváltoztatása mindig elronthat bármilyen
szimmetriára vonatkozó felvetést. Ez az anyagi paraméter ráadásul nem határozható
meg csupán mechanikai mérésekkel, mert a belső változóra csak következményeiben,
\re{ptj} paramétereinek meghatározásából szerezhetnénk információt. 
Ha pedig \(\bar G\) értékét egységnek választjuk és csak szimmetrikus, vagy
antiszimmetrikus vezetési együtthatókat feltételezünk, akkor szembesülünk
azzal, hogy modellünk nem jó a tapasztalatok egy részére (\cite{Ver97b} p98). Másrészt
viszont csak a vezetési mátrix szimmetrikus része jelent disszipációt, és annak 
antiszimmetrikus - "giroszkópikus" - része nem
ad járulékot az entrópiaprodukcióhoz. Azaz,  általános vezetési mátrixok
használata  lehetőséget
teremt a termo\-dinamikai leírás  érvényességi körének jelentős kiterjesztésére
is. Erre vonatkozóan fontos példát jelent, hogy a \nev{Maugin}-féle dinamikai 
szabadsági fokok és a belső változók elmélete
csak az általános esetben egyesíthető \cite{VanAta08a}.  Megjegyezzük, hogy
az \nev{Onsager} által adott bizonyítás a reciprocitási relációkra csak tiszta
mikroszkopikus háttér esetén érvényes, ezért fenti megállapításunk nincs
ellentmondásban vele.
 
 A vezetési együtthatók mátrixának szimmetriája az úgynevezett disszipációs potenciálok létezésének
is szükséges és elégséges feltétele. A disszipációs potenciál
a termodinamikai erőknek (vagy áramoknak) olyan függvénye, amelynek parciális
deriváltjaiként kapható meg a fenti  (\ref{lo1})--(\ref{lo2}) vezetési egyenletek 
jobb oldala. Jelen esetben, ha $l_{12} =l_{21}$, akkor
\begin{equation}
\Phi(\dot\ep, -\bar G\xi)= l_1 \frac{\dot\ep^2}{2} - l_{12}\dot{\ep}\bar G\xi + l_2\frac{(\bar G\xi^2)}{2},
\end{equation}
hiszen
\begin{eqnarray*}
\pd{\Phi}{\dot{\ep}} &=& l_{1} \dot\ep-  l_{12} \bar G\xi, \\
\pd{\Phi}{(-\bar G\xi)} &=& l_{12} \dot\ep-  l_{2} \bar G\xi. 
\end{eqnarray*}

A disszipációs potenciálok a kapcsolt képlékenység termodinamikai elméletében kulcsfontosságúak,
 a
folyásfüggvény, illetve a képlékeny potenciál sze\-repét játsszák.

\section{ képlékenység termodinamikai elmélete - kis deformációk}

 A termodinamikai képlékenység is egy belső változós elmélet, ahol a 
belső változót azonnal fizikai,
kinematikai jelentéssel felruházva, képlékeny deformációként vezetik be.
Egy mechanikai leírásban nincs is sok más választásunk, mechanikai fogalmakkal kell
megragadnunk a jelenségeket. A képlékenység  oka valamilyen belső szerkezeti változás az anyagban (pl.
diszlokációk mozgása, de a szemcsék átrendeződése is ide tartozhat), amely
megváltoztatja a mechanikai erőket, az anyag belső feszültségviszonyait. Azaz, a jelentkező
deformáció csak
következmény, nem világos, hogy milyen feltételekkel jelentheti alapját a modellezésnek.
Ráadásul a fizikai kép, a változó kinematikai jelentése is  lényeges,   ennek
felületes kezelése következetlenné teszi a képlékenységelméleteket, ahogy ezt a nagy alakváltozások
esetére   Bertram megmutatta \cite{Ber05b}. Egy termodinamikai leírásban -
általáno\-sabb alapokon, a  releváns fizikai mennyiséget belső változóként bevezetve -  vizsgálhatóbb a képlékenységre
vezető feltételrendszer.

Ez a belső változó azonban reológiai társától eltérően nem tekinthető egyúttal
\nev{Verhás}-féle dinamikai szabadsági foknak (a \nev{Verhás}-féle dinamikai szabadsági
fok olyan speciális belső változó, amely termodinamikai egyensúlyban nulla,
azaz a hozzá tartozó intenzív mennyiség arányos magával a változóval). Ugyanis
a szóban forgó szerkezeti változások maradandóak lehetnek, a külső hatás megszűnésekor
sem enyésznek el. Éppen
ezért a képlékenység tárgyalásakor már sztatikai szinten is az előző fejezet
reológiai modelljétől eltérő feltevéseket teszünk a belső változó és a deformáció viszonyára, elképzelve,
hogy a belső változó változtatja, adott feltételekkel csökkenti a feszültséget, illetve végső soron  a tárolt
rugalmas energiát. Ebből következően
az $e_B$ belső e\-nergia is különbözik a reológiai tárgyalásban bevezetett
\re{rben} formától: 
\begin{equation}
  e_B= e- \frac{v^2}{2}- e_{p}(\ep^{ij},\xi^{{ij}}).
\label{kben}\end{equation}
 
A sztatikus mechanikai feszültséget az előző rugalmas-reológiai esethez
hasonlóan, az \(e_{p}\) energia deformáció szerinti deriváltjaként határozzuk meg: 
\begin{equation}
(t_s)^{ij}=\pd{e_{p}}{\ep_{ij}}.
\label{sfesz}\end{equation}

Feltételezzük, hogy a belső változó megváltozása a deformáció változásához képest
ellentétesen hat a feszültségre. A deformáció hatását a feszültségre a rugalmassági
modulussal jellemezzük,  azaz nemlineáris esetben a feszültségnek   a deformáció szerinti
deriváltjával. A feszültségnek a belső változó szerinti deriváltja
hasonló negyedrendű tenzor lesz, ezt a képlékeny deformációhoz tartozó rugalmassági
modulusnak tekintjük. A fenti követelmény alapján a kétfajta rugalmassági tenzor arányos kell legyen,
méghozzá negatív együtthatóval, azaz
\begin{equation}
 \pd{t^{ij}}{\ep_{kl}} + A^{kl}_{mn}  \pd{t^{ij}}{\xi_{mn}} =0,
\label{kdef}\end{equation}
ahol $ A^{kl}_{mn}$ szimmetrikus és pozitív definit abban az értelemben,
hogy   $ A^{kl}_{mn}= A_{kl}^{mn}$, továbbá  $z_{ij} A^{ij}_{kl}
z^{kl}\geq 0$ minden  $z_{ij}\neq 0$-re. Ezenkívül $ A^{kl}_{mn}= A_{nm}^{kl}$ és
$ A^{kl}_{mn}= A_{mn}^{lk}$ a deformáció és a belső változó szimmetrikussága miatt.
Ha  $ A^{kl}_{mn}$
állandó, akkor a \re{kdef} feltétel úgy is felfogható, hogy a deformációval
arányos a belső változó, azaz 
\begin{displaymath} 
  \ep^{kl} = A^{kl}_{mn} \xi^{mn}, 
\end{displaymath}
hiszen \re{kdef} megoldása alapján  \(t^{ij}(\ep^{kl}-A^{kl}_{mn} \xi^{mn})\).
Ezért, a sztatikus feszültség előbbi \re{sfesz} definíciója szerint az \(e_{pla}\)  rugalmas-képlékeny energia
\begin{equation*}
   e_{pla}(\ep^{ij},\xi^{ij}) = e_{rug}(\ep^{ij}-A^{ij}_{mn} \xi^{mn})+e_{din}(\xi^{ij}).
\end{equation*}
Vagyis a reológiai esethez teljesen hasonló formát kaptunk azzal a különbséggel,
hogy a deformálódást a belső változó növekedése csökkentheti. Másként
fogalmazva, a rugalmas energia csak a deformáció rugalmas részétől függ, amely a 
valódi és a képlékeny
deformáció különbségeként áll elő. Bevezethetünk tehát egy új változót:
\begin{equation*}
  (\ep_p)^{ij}=A^{ij}_{mn} \xi^{mn}.
\end{equation*}
Ez a változó pedig már képlékeny deformációként értelmezhető, visszakapjuk
\re{addpla}-et. Segítségével a fenti 
rugalmas-képlékeny energia formája
\begin{equation}
   e_{pla}(\ep^{ij},(\ep_p)^{ij}) = e_{rug}(\ep^{ij}-(\ep_p)^{ij})+e_{din}((\ep_p)^{ij}).
\label{ken}\end{equation}
A továbbiakban feltételezzük, hogy mind a rugalmas, mind a dinamikai energia
kvadratikus és izotróp függvénye változóinak, azaz
\begin{eqnarray}
e_{rug}(\ep^{ij}-(\ep_p)^{ij}) &=&  
        \frac{\lambda}{2} \left(\ep^{i}_i-(\ep_p)^{i}_i\right)^2 + 
        \mu \left(\tilde\ep^{ij}-(\tilde\ep_p)^{ij}\right)
        \left( \tilde\ep_{ij}-(\tilde\ep_p)_{ij}\right),\\ 
e_{din}((\ep_p)^{ij}) &=&  
        \frac{\bar\lambda}{2} \left((\ep_p)^{i}_i\right)^2 + 
        \bar\mu (\tilde\ep_p)^{ij}(\tilde\ep_p)_{ij}.
\end{eqnarray} 
 
Itt hullámmal az adott szimmetrikus másodrendű\ tenzor nulla nyomú részét
jelöltük az eddigiekhez hasonlóan, \(\lambda\) és  \(\mu\) a rugalmas \nev{Lamé}-állandók,  
\(\bar\lambda, \bar\mu\) pedig a képlékeny (keményedő) tartományban
érvényes analóg anyagi paraméterek. 
  
A termodinamikai leírás ezek után is teljesen analóg a reológiai esettel. Az
energia járulékainak azonosítása után az entrópia csak a belső energián keresztül 
függ az összdeformációtól és a képlékeny deformációtól: 
\(s(e_{b},\epsilon^{ij},(\ep_p)^{ij})=\hat s(e_B)\). Ennek megfelelően a \nev{Gibbs}-reláció 
a következő:
\begin{equation}
 de_B  = Td\hat s+(t_s)_{ij}d\epsilon^{ij}+ (t_p)_{ij} d(\ep_p)^{ij},
\end{equation}  
ahol  $(t_p)^{ij}=\pd{e_{pla}}{(\ep_p)_{ij}}$ a képlékeny feszültség.
Az eddigieket összefoglalva az intenzív mennyiségeket a fenti \nev{Gibbs}-reláció,
 illetve \re{kben} és \re{ken} alapján
a következőképpen adhatjuk meg:  
\begin{eqnarray*}
\pd{\hat s}{e_B} =\pd{s}{e_b}&=& \frac{1}{T}, \\
\pd{s}{\epsilon^{ij}} &=&
        -\frac{1}{T}\pd{e_{pla}}{\epsilon^{ij}}=
        -\frac{1}{T}\pd{e_{rug}}{\epsilon^{ij}} =
        -\frac{(t_s)^{ij}}{ T},\\
\pd{s}{(\ep_p)^{ij}} &=&
        -\frac{1}{T}\pd{e_{pla}}{(\ep_p)^{ij}}=
        -\frac{(t_p)_{ij}}{T}.
\end{eqnarray*}

Ha a deformáció nem nulla, akkor  belső változónk értéke termodinamikai egyensúlyban
nem feltétlen nulla. Ezért a belső változó nem lesz {Verhás-féle dinamikai szabadsági fok},
ellentétben a reológiai esettel. Az előző fejezet számításait megismételve
végülis az entrópiaprodukció következő lesz:
 
\begin{equation}
 T\sigma_{s}= \left(t^{ij}-(t_s)^{ij}\right)\dot\epsilon_{ij}-(t_p)^{ij} (\dot\ep_p)_{ij}
    \geqslant 0.
\label{epk}\end{equation} 

Ezért aztán, szem előtt tartva, hogy  mind a sztatikus, mind a képlékeny
fe\-szültség ismert függvénye az alapváltozóknak, a megfelelő termodinamikai erők és áramok a
következőek lesznek: 
\begin{center}
\begin{tabular}{c|c|c}
Erő & $\dot\ep^{ij}$ & $-(t_p)^{ij} $ \\\hline
Áram & $t^{ij}-(t_s)^{ij}$ & $(\dot\ep_p)_{ij}$ \\
\end{tabular}
\end{center}

A lineáris vezetési egyenletek ezek után 
\begin{eqnarray}
t^{ij}-(t_s)^{ij} &=& L_{11}^{ijkl} \dot\ep_{kl}-  L_{12}^{ijkl} (t_p)_{kl} ,\label{ko1}\\
(\dot\ep_p)^{ij} &=&  L_{21}^{ijkl}  \dot\ep_{kl} -  L_{22}^{ijkl} (t_p)_{kl}.
\label{ko2}\end{eqnarray}

A fenti egyenletrendszer izotrop esetben érvényes változata is nagyon hasonló,
mint az előző fejezetben:
\begin{eqnarray}
t^i_{\;i}-(t_s)^i_{\;i} &=& 
        m_{11} \dot\ep^i_{\;i} - m_{12} (t_p)^{i}_i,\label{kio1}\\
(\dot\ep_p)^i_{\;i} &=&  
        m_{21} \dot\ep^i_{\;i}  -  m_{22} (t_p)^i_{\;i},\label{kio2}\\
\tilde t^{ij}-(\tilde t_s)^{ij} &=& 
        k_{11}\dot{\tilde  \ep}^{ij}-  k_{12}(\tilde t_p)^{ij},\label{kio3}\\
(\dot{\tilde\ep}_p)^{ij} &=&  
        k_{21}\dot{\tilde  \ep}^{ij} -  k_{22}(\tilde t_p)^{ij}.
\label{kio4}\end{eqnarray}

A belső változók kiküszöbölése most  is lehetséges, ha feltételezzük, hogy
a vezetési együtthatók állandóak. 

Ez az egyenletrendszer azonban még nem
képlékenységi elmélet, a képlékeny viselkedés egy fontos eleme, maga a képlékenységi
feltétel és határ sehol sem jelenik meg benne. A megoldásai is mutatják, hogy
eddigi feltételeink sem tartalmazzák rejtetten. A képlékeny deformációnak ugyanis megvan az a tulajdonsága, amit a bevezetésben
említett összes elmélet alapul vesz, hogy csak bizonyos feszültség-, energia-,
vagy egyéb feltétel teljesülése esetén kezd változni. A viselkedés a tapadási
súrlódáshoz hasonló, a termodinamikai képlékenység elméletének mechanizmusát
alkalmaztuk a második fejezetben. Ha a képlékenység csak deviatorikus feszültség 
hatására lép fel, mint a fémekben, akkor
az ott leírt gondolatmenet alapján reológiai motivációjú egyenletrendszerünk
módosítása kézenfekvő. Legyen \re{kio4}-ben a $k_{22}$ együttható nemlineáris
és  \re{onsmi} szerint függjön a termodinamikai erőtől, jelen esetben $(t_p)^{ij}$-től. Természetesen
a kontinuummodell bonyolultabb egyenletei több kérdést nyitva hagynak, de
az egyik legegyszerűbb olyan változtatás, amelytől kép\-lékeny viselkedést várhatunk,
a következő:
\begin{equation}
k_{22} = \frac{ k_2 }{1+ \frac{k_2}{\sigma_c} |\tilde t_p|}. 
\end{equation} 

Itt $|\tilde t_p| = \sqrt{(\tilde t_p)^{ij}(\tilde t_p)_{ij}}$ a képlékeny feszültség deviatorikus
részének abszolút
értéke. Ez a kép\-lékenységi modellünk kulcsfontosságú utolsó feltevése.

A kapott vezetési egyenletekkel az entrópiaprodukció 
\begin{multline}
 T\sigma_{s}=m _{11}(\dot\epsilon_{i}^i)^2-
        (m_{12}+m_{21})\dot\ep^i_i(t_p)^i_i+
        m_{22}(t_p)^{i2}_i+\\
        k _{11}\dot{\tilde\epsilon}_{ij}\dot{\tilde\epsilon}^{ji}-
        (k_{12}+k_{21})\dot{\tilde\epsilon}_{ij}(\tilde t_p)^{ij} +
         \frac{ k_2 }{1+ \frac{k_2}{\sigma_c} |\tilde t_p|}
         |\tilde t_p|^2 
    \geqslant 0.
\end{multline} 
  
Az utolsó tag figyelemre méltó. Ugyanis, ha $k_2 |\tilde t_p|/\sigma_c\gg 1$
akkor a $\sigma_c |\tilde t_p|$ formára egyszerűsödik, ekkor a többivel ellentétben
nem kvadratikus. A következő fejezetben látni fogjuk, hogy ez felel meg az
ideális képlékenységnek. A termodinamikai áram abszolút értékét tartalmazó entrópiaprodukció
--- a kvadratikus forma helyett --- a termodinamikai képlékenységelmélet védjegyszerű jellemzője.
Ha a fenti vezetési mátrix szimmetrikus, akkor az entrópiaprodukcióból könnyen megadhatjuk a 
vonatkozó disszipációs potenciált, amit az első fejezetben mondottak alapján
a képlékeny potenciállal azonosíthatunk:
\begin{multline}
 \Phi\left(\dot\ep^i_i,\dot{\tilde\ep}^{ij},(t_p)^{i}_i,(\tilde t_p)^{ij}\right) =
        \frac{m _{11}}{2}(\dot\epsilon_{i}^i)^2-
        m_{12}\dot\ep^i_i(t_p)^i_i+
        \frac{m _{22}}{2}(t_p)^{i2}_i+\\
        \frac{k _{11}}{2}\dot{\tilde\epsilon}_{ij}\dot{\tilde\epsilon}^{ji}-
        k_{12}\dot{\tilde\epsilon}_{ij}(\tilde t_p)^{ij} +
        \sigma_c |\tilde t_p| -
        \frac{\sigma_c^2}{k_2}\ln\left(1+ \frac{k_2}{\sigma_c} |\tilde t_p|\right)
\end{multline} 
 
A disszipációs potenciál egyes változói szerinti deriváltak a \re{kio1}--\re{kio4}
egyenletek jobb oldalai adják. 

A továbbiakban néhány nagyon egyszerű esetben szemléltetni fogjuk, hogy
valóban képlékenységi elméletet adtunk meg, méghozzá egy dinamikus, kinematikai 
keményedő és disszipatív képlékenységi elméletet, a viszkoelasztoplaszticitás
talán legegyszerűbb modelljét.

\subsection{Közönséges képlékenység - képlékenység homogén testekre}

Hasonlóan a reológiai esethez, a homogén képlékeny test egyenletei is többféle terhelés
hatására jöhetnek létre. 
 
A rugalmas-képlékeny energia kvadratikus formája, azaz \re{ken} az egydimenziós
homogén esetben a következő lesz:
\begin{equation}
e_{pla} (\ep, \ep_p) =
        G\frac{(\varepsilon-\ep_p)^2}{2}+ \bar G\frac{\ep_p^2}{2}.
\label{plaer}\end{equation} 
Ennek megfelelően a sztatikus  feszültség és képlékeny feszültség \begin{eqnarray*}
  \pd{ e_{pla}}{\ep} &=&    G(\varepsilon-\ep_p),\\ 
 \pd{ e_{pla}}{\ep_p} &=& G(\ep_p-\varepsilon)+\bar G\ep_p = 
        -G\varepsilon+(G+\bar G)\ep_p.
\end{eqnarray*}

A rugalmas-képlékeny energiafüggvény konvex, ha $G$ és $\bar G$ pozitív. Emlékeztetünk, hogy ez
a termosztatikai
kép interpretálja belső változónkat képlékeny deformációként, a két állandót
pedig a rugalmas és a keményedési   modulusként (tehát a $\bar G =0$ esetben
lesz a képlékenység ideális). 

Az entrópiaprodukció \re{epk} formulája változatlan marad:
\begin{equation}
 T\sigma_{s}= \left(t - \pd{ e_{pla}}{\ep}\right)\dot\epsilon-
       \pd{ e_{pla}}{\ep_p}\dot\ep_p \geqslant 0.
\label{plas}\end{equation}  

A termodinamikai erők és áramok pedig a termosztatikai és a képlékeny feszültségekre
bevezetett \(t_s=\pd{ e_{pla}}{\ep}\) és \(t_p=\pd{ e_{pla}}{\ep_p}\)
jelölésekkel

\begin{center}
\begin{tabular}{c|c|c}
Erő & $\dot\ep$ & $-t_p$ \\\hline
Áram & $t- t_s$ & $\dot\ep_p$ \\
\end{tabular}.
\end{center}
A vezetési egyenletek ezek után a következőek
\begin{eqnarray}
 (t- t_s)   &=& l_{1} \dot\ep - l_{12} t_p, \\
\dot \ep_p  &=& l_{21} \dot\ep - l_{2}t_p.
\end{eqnarray}
Ezeket most átranszformáljuk az úgynevezett
ve\-gyes erő-áram reprezentációba \cite{Gya67b}. Az átrendezés
a szigorúan lineáris esetben (konstans és invertálható vezetési mátrix) ekvivalens az előző
fejezet reprezentációjával \cite{Mei73a,Gya70b}:
\begin{eqnarray}
\dot \varepsilon=l^{-1}_{1} (t- t_s) - 
         l^{-1}_{1}l_{12} t_p  &=&
          \hat l_{1}  (t- t_s) + 
        \hat l_{12} t_p, \\
\dot \ep_p =  l^{-1}_{1}l_{21} (t- t_s) + 
        (-l_{2}+ l^{-1}_{1}l_{12}l_{21})t_p  &=&
         \hat l_{21} (t- t_s) + 
        \hat l_{2}t_p.
\end{eqnarray}

A teljes nemlineáris transzformáció helyett viszont a nemlinearitást most más, a képlékenységi irodalomban hagyományos módon,
\re{onsm1}-el analóg formában fogjuk a számításokban használni.  Tegyük fel
tehát, hogy az \(\hat l_{2}\) együttható speciális. Egyrészt tartalmaz
egy konstans, reológiai tagot, amely (\ref{ptj}) szerint a relaxációs és disszipatív
hatásokért lesz felelős. Másrészt, ha ez a konstans tag nulla, akkor a képlékeny deformációhoz
tartozó termodinamikai intenzív paraméter $t_{p}=\pd{ e_{pla}}{\ep_p} $ - jelen esetben
egyúttal termodinamikai erő - csak a kép\-lékeny deformáció \textit{előjelét}
képes meghatározni, nagyságát nem.  Azaz  feltételezzük, hogy
\begin{equation}
\hat l_2 =|\dot \ep_p|/\sigma_c+ l,
\label{hpq}\end{equation}
ahol \(l\) és \(\sigma_c\) pozitív állandók. Látni fogjuk, hogy $\sigma_c$  a
folyási 
határfeszültség szerepét játsza. 
Az \(\hat l_{1}, \hat l_{12}, \hat l_{21}, l\) vezetési együtthatókra a nemnegatív
entrópiaprodukcióból következő szokásos termodinamikai egyenlőtlenségek érvényesek.
Látni fogjuk, hogy ez a fajta - termodinamikailag következetlen, de a képlékenységtanban
szokásos - nemlinea\-ritás ugyanolyan hatást eredményez, mint
amit a tapadási súrlódás kapcsán már tapasztaltunk: a képlékeny deformáció
csak egy feszültségküszöb átlépése után kezd növekedni.
  
Az entrópiaprodukció, illetve az energia disszipáció megfelelő tagja most sem kvadratikus, hanem \(|\dot\ep_p|\)-vel arányos, ha $l=0$.
Ez a vegyes reprezentáció miatt nem egészen nyilvánvaló, mert ha \re{plas}-be egyszerűen visszahelyettesítjük a fenti
vezetési egyenleteket, kvadratikus formát kapunk. Teljes áramreprezentációt választva, 
azaz a termodinamikai erőkkel kifejezve a termodinamikai
áramokkal és behelyettesítve az entrópiaprodukcióba nem keveredik az $\hat
l_2 $ vezetési
együttható $\dot\ep_p $ függése a hozzá tartozó termodinamikai erővel. Ez a következmény
- vagyis, hogy a disszipáció a képlékeny deformáció időderiváltjának elsőrendű
homogén függvénye -
a termodinamikai képlékenység elméletének kiindulópontja szokott lenni. 

A
vezetési együtthatók tulajdonságait kidomborító tárgyalásunk arra mutat
rá, hogy itt a második főtétel egyenlőtlenségének egy olyan megoldásáról
van szó, a\-mely az irreverzibilitás egy,  a megszokottól eltérő, új módját reprezentálja.
A lineáris vezetési együtt\-hatók a súrlódásos, diffúziós, relaxációs jelenségekben
megnyilvánuló disszipációt jellemzik, a tapadási súrlódásos jellegűek pedig  hiszterézises  jelenségekben jelentkező irreverzibilitás
 mechanizmusát mutatják meg.

 A képlékenység  elméletének alapfeltevése, hogy a képlékeny
deformáció csak egy bizonyos feszültség felett lép  fel. A termodinamikai
elmélet szépsége, hogy ezt a viselkedést a klasszikus képlékenységnél mélyebb
szinten modellezi, mivel nem közvetlenül a tapasztalt következményeket, hanem az okokat próbálja matematikailag
megragadni. 

\subsection{Megoldások}

A megoldandó differenciálegyenlet-rendszer tehát a fentiek alapján az időderiváltak
kiküszöbölésével adódik:

Ha $\dot \ep_p>0$, akkor 
\begin{eqnarray}
\dot \varepsilon &=&  \hat l_1 \left(t- \pd{ e_{pla}}{\ep}\right) + 
        \hat l_{12} \pd{ e_{pla}}{\ep_p}  = 
         \hat l_1 (t - G(\ep-\ep_p)) + 
        \hat l_{12} (-G\ep+(G+\bar G)\ep_p), \label{omHP}\\
\dot \ep_p &=& \frac{\hat l_{12} \left(t- \pd{ e_{pla}}{\ep}\right) + 
       l\pd{ e_{pla}}{\ep_p}}{1+\sigma^{-1}_c\pd{ e_{pla}}{\ep_p}} 
      = -\sigma_c{l} + \frac{\hat l_{12} (t - G(\varepsilon-\ep_p)) + 
      \sigma_{c}l}{1 + \sigma_{c}^{-1}(-G\varepsilon+(G+\bar G)\ep_p)} 
        \label{oxHP}.
\end{eqnarray}
  
Ha $\dot \ep_p\leq 0$, akkor  
\begin{eqnarray}
\dot \varepsilon &=&  \hat l_1 \left(t- \pd{ e_{pla}}{\ep}\right) +
        \hat l_{12} \pd{ e_{pla}}{\ep_p}\label{omHP1} = 
         \hat l_1 (t - G(\varepsilon-\ep_p)) + 
        \hat l_{12} (G\ep-(G+\bar G )\ep_p), \\
\dot \ep_p &=& \frac{\hat l_{12} \left(t- \pd{ e_{pla}}{\ep}\right) + 
       l\pd{ e_{pla}}{\ep_p}}{1-\sigma_{c}^{-1} \pd{ e_{pla}}{\ep_p}} 
      = {\sigma_k}l + \frac{\hat l_{12} (t - G(\ep-\ep_p))- 
      \sigma_c l}{1-\sigma^{-1}_c (-G\varepsilon+(G+\bar G)\ep_p)} 
        \label{oxHP1}.
\end{eqnarray}

Vagyis, attól függően, hogy a képlékeny deformáció növekszik, vagy csökken,
a $\sigma_c$ előtti előjelet megváltozatjuk az \re{omHP}-\re{oxHP} egyenletrendszerben.
Az itt tárgyalt homogén esetben  a \nev{Poynting-Thomson}-modellhez hasonlóan \(\ep_p\)
akár ki is kiküszöbölhető. A (\ref{omHP})-(\ref{oxHP1}) egyenletekkel egy reológiai-képlékeny
\nev{Poynting-Thomson}-testet adtunk meg  feszültségi (\nev{Tresca}) típusú képlékenységi  feltétellel.

\begin{figure}
\includegraphics[width=0.8\textwidth]{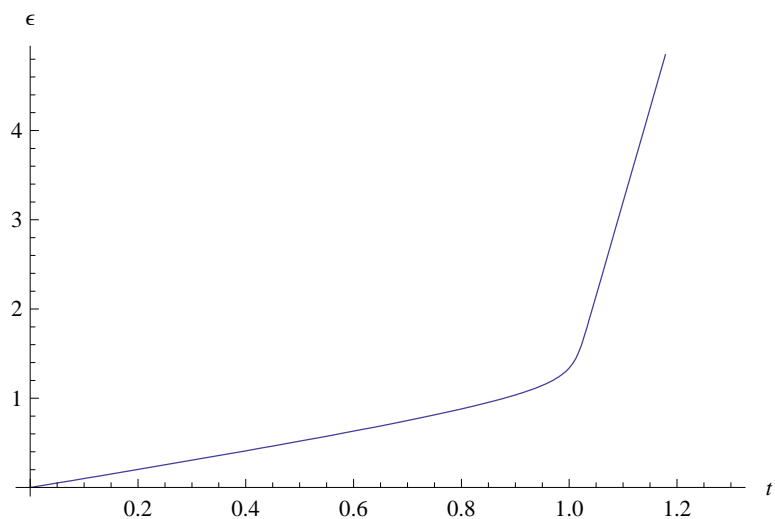}
\caption{\label{Fig1}
Mechanikai egyensúly, a deformáció időfüggése}
\end{figure}

\begin{figure}
\includegraphics[width=0.8\textwidth]{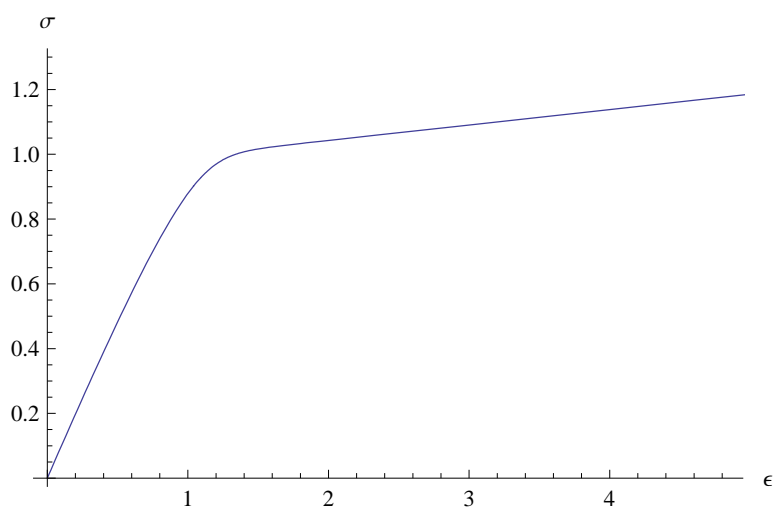}
\caption{\label{Fig2}
Mechanikai egyensúly, a feszültség deformációfüggése}
\end{figure}

\begin{figure}
\includegraphics[width=0.8\textwidth]{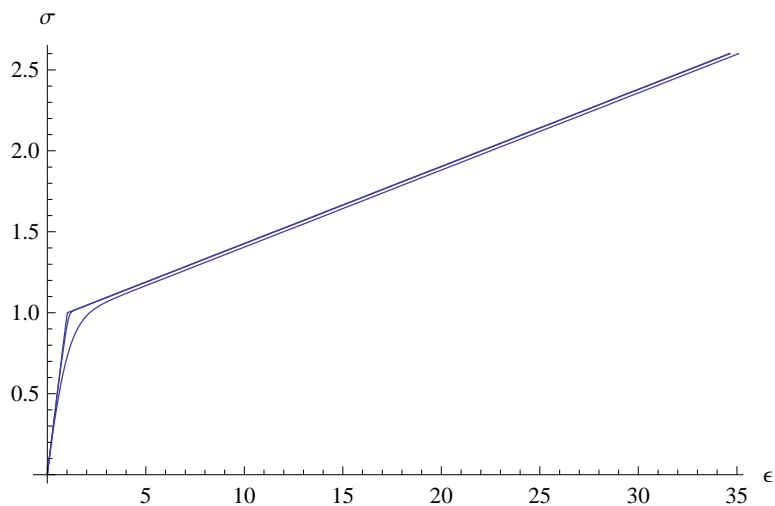}
\caption{\label{Fig3}
Az $l$ paraméter szerepe ($l = 0.01, 0.1, 1$, felülről lefele).}
\end{figure}

Tekintsünk először egy mechanikai egyensúlyi esetet, amikor nincs viszkozitás,
a feszültség megegyezik a termosztatikai feszültséggel, azaz $t= t_s=\pd{ e_{pla}}{\ep}$, a viszkózus feszültség nulla. Legyen a felterhelés
 sebessége \(v=1\), a határfeszültség  \(\sigma_c = 1\), a további paraméterek
 értékei pedig  \(l = 0.05\),  \(G = 1\) és \( \bar G = 0.05\) \(\epsilon(0)
 = 0\). Ekkor a deformáció időfüggése az \ref{Fig1}. ábrán, a feszültség
 deformációfüggése pedig a \ref{Fig2}. ábrán látható. Az \(l\) képlékenységi
 paraméter szerepét a \ref{Fig3} ábrán szemléltetjük, ahol \(l=0.01,0.1,1\),.
Itt a kisebb paraméter élesebb folyáshatárt jelent. \(l\) és ${\hat l_1}$ szerepe
 tulajdonképpen hasonló, ha együtt lépnek fel. Nagy ${\hat l_1}$ és kis \(l\) eredményez
 éles képlékeny átmenetet.

\begin{figure}
\includegraphics[width=0.8\textwidth]{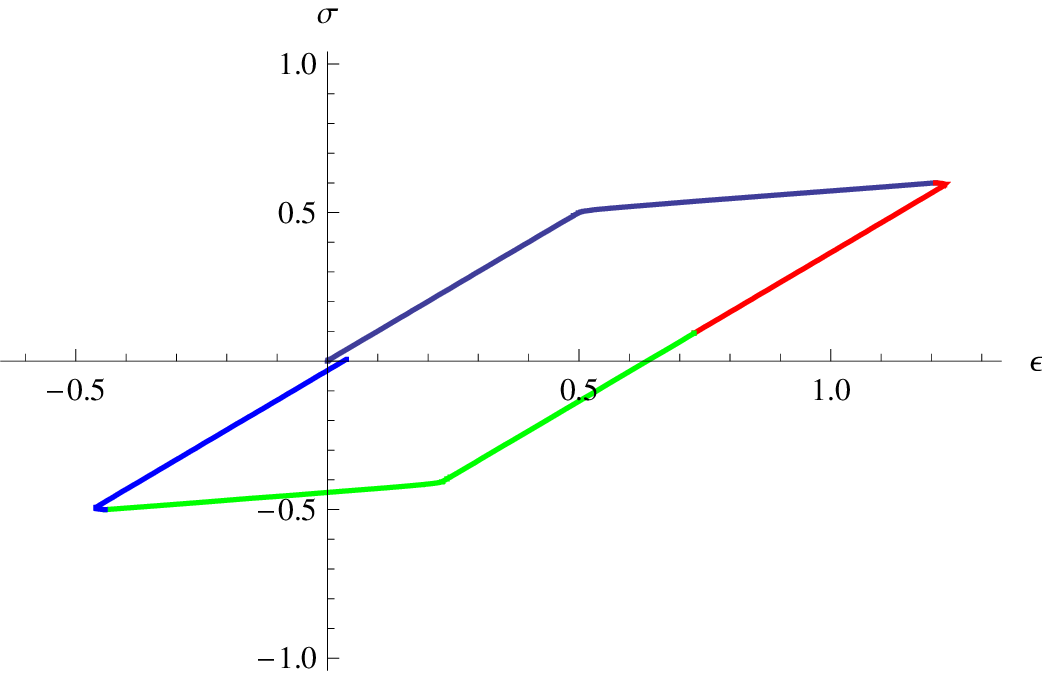}
\caption{\label{Fig4}
Hiszterézis}
\end{figure}


A fenti egyenletrendszer egy lineárisan kinematikai keményedési modell, ahogy azt a
4. ábra mutatja. Itt \(\sigma_c=0.5\) és \(t=0.6\)-nál a terhelési 
sebesség előjelet vált. A színváltások a képlékeny deformáció monoton változásának
szakaszait jelzik, amikor a differenciálegyenletben a megfelelő tag előjelet
vált. Megfigyelhető a "ratcheting" jelensége is, bár itt csak egyetlen ciklust
ábrázoltunk.

\section{Következtetések és megjegyzések}

A képlékenységtanba disszipációt is bevezető elméletek általában csak a \nev{Kelvin-Voigt}-testet
tartalmazzák,
azaz csak a viszkózus hatásokat képesek modellezni. Maugin ugyan említi, hogy a relaxációs
hatásért felelős \nev{Maxwell}-test
is a termodinamikai elmélet része, de annak modellezésére külön belső változót
vezet be \cite{Mau99b}. Másrészt, a keményedésért és a képlékenységért felelős
belső változókat is sokan különválasztják \cite{JirBaz02b}. Az általunk tárgyalt
modellben egyetlen belső változó modellezi a kúszási, a relaxációs és a képlékenyedési
jelenségeket. 
  
A képlékenységre vezető alapfeltevés az, hogy a vezetési egyenletek által
eredményezett disszipáció nem feltétlen a klasszikus lineáris és konstans együtthatók
által sugallt kvadratikus, a termodinamikai áramokban másodrendűen homogén
formájú lehet, hanem  attól eltérő, a termodinamikai áramokban
elsőrendűen pozitívan homogén is.
Speciálisan a belső változóhoz tartozó tagról elegendő feltételezni ezt a
tulajdonságot. Az általunk itt javasolt \nev{Onsager}-együttható
formák erre vezetnek.

A hagyományos tárgyalásban alapfogalomként --- képlékeny potenciálként és egyúttal
folyásfüggvényként  - használt
disszipációs potenciálok létezé\-sének
feltétele a \nev{Onsager} reciprocitási relációinak fennállása, azaz a szimmetrikus vezetési
mátrix (pontosabban a nemlineáris \nev{Gyarmati-Li}
reciprocitási relációkat kell megkövetelnünk  \cite{Gya61a,Li62a}). Ezért egy általános tárgyalásban a belső
változó időderiváltjától függő disszipációs potenciál 
helyett érde\-mes a vezetési egyenletekből kiindulni. Ez esetben vizsgálható, hogy ez az
általánosítás mennyiben vezet a nem kapcsolt képlékenység leírásra. 
  
Viszkózus, reológiai hatások regularizálják és stabilizálják, megoldhatóbbá teszik az  az ideális képlékenység
egyenleteit  \cite{Gur00a}. Az egyenletek  (teljes parciális differenciálegyenlete rendszer)
sajátos szerkezete miatt a szokásos numerikus stabilizálási technikák (hiperbolikus 
kiegészítés, numerikus viszkozitás) nem működnek a kép\-lékenység esetén. A 
fenti differenciálegyenletek például \(l=0\) esetben már nem differenciálegyenletek,
ezért nehezebben  értelmezhetőek és tárgyalhatóak numerikusan.
Tárgyalásmódunk értelmezi a képlékeny deformációval kapcsolatos irreverzibilitást,
ezért klasszikus képlékenységelméleten túlmutató, valódi dinamikus képlékenységi feladatok esetén lehet
jelentősége, mint például a képlékenyedési frontok 
terjedésének modellezése.

\section{Köszönetmondás} 

Köszönet \nev{Matolcsi Tamás}nak, aki rávilágított a képlékenységelmélet működésére,
\nev{Asszonyi Csabá}nak,  \nev{Fülöp Tamás}nak és \nev{Fekete Tamás}nak akikkel együtt
még most sem gondolom,
hogy ez ilyen egyszerű lenne, de akik bíznak benne, hogy már így is sok mindenre
jó. A munkát az Otka  K81161 pályázatával támogatta.

\bibliographystyle{unsrt}

\end{document}